\documentstyle[aps,prb,floats,epsfig,psfig,12pt]{revtex}
\def\beq{\begin{equation}}
\def\cP{{\cal P}}
\def\cf{{\it cf.}}
\def\diag{\mbox{diag}}
\def\eeq{\end{equation}}
\def\eg{{\it e.g.}}
\def\H{{\cal H}}
\def\hLam{\hat\Lambda}
\def\hN{\hat N}
\def\ie{{\it i.e.}}
\def\k{{\rm\bf k}}
\def\blambda{\bar\lambda}
\def\lambdac{\lambda_{\rm c}}
\def\M{M_{\rm c}}
\def\MC{Monte Carlo}
\def\P{{\hat P}}
\def\hcP{{\hat\cP}}
\def\hP{{\hat P}}
\def\psiT{{\psi_{\rm T}}}
\def\psiB{{\psi_{\rm B}}}
\def\ket#1{| {#1} \rangle}
\def\psiTket{\ket{\psiT}}
\def\br#1{ \langle{#1}}
\def\bra#1{ \langle{#1}|}
\def\psiTbra{\bra{\psiT}}
\def\psiTbr{\br{\psiT}}
\def\psit#1{\psi_{{\rm T}#1}}
\def\psitbra#1{\bra{\psit #1}}
\def\psitbr#1{\br{\psit #1}}
\def\psitket#1{\ket{\psit #1}}
\def\tN{\tilde{N}}
\def\tP{\tilde{{\cP}}}
\def\tlambda{\tilde\lambda}
\def\tpsi{{\tilde{\psi}}}
\def\tPsi{{\tilde{\Psi}}}
\def\viz{{\it viz.}}
\def\vs{{\it vs.}}
\begin{document}\tightenlines
\title{Monte Carlo
  computation of correlation times of independent relaxation modes at
  criticality} 
\author{M. P. Nightingale} 
\address{Department of Physics, University
  of Rhode Island, Kingston RI 02881, USA} 
\author{H.W.J. Bl\"ote }
\address{Faculty of Applied Sciences, Delft University, P.O. Box 5046,\\
  2600 GA Delft, The Netherlands;} 
\address{Lorentz Institute, Leiden University, Niels Bohrweg 2,\\
  P.O. Box 9506, 2300 RA Leiden, The Netherlands} \maketitle

\begin{abstract}
  We investigate aspects of universality of Glauber critical
  dynamics in two dimensions. We compute the critical exponent $z$ and
  numerically corroborate its universality for three different models in
  the static Ising
  universality class and for five independent relaxation modes.  We
  also present evidence for universality of amplitude ratios, which
  shows that, as far as dynamic behavior is concerned, each model in a
  given universality class is characterized by a single non-universal
  metric factor which determines the overall time scale.  This paper
  also discusses in detail the variational and projection methods that
  are used to compute relaxation times with high accuracy.
  \vfill
\noindent
pacs {64.60.Ht, 02.70.Lq, 05.70.Jk, 64.60.Fr}\\
\end{abstract}

\section{introduction}

Critical-point behavior is a manifestation of power-law divergences of
the correlation length and the correlation time. The power laws that
describe the divergence of the correlation length on approach of the
critical point are expressed by means of critical exponents that are
dependent on the {\em direction} of this approach, which may, \eg, be
ordering-field-like or temperature-like. The exponents describing the
singularities in {\it thermodynamic} quantities can be expressed in
terms of the same exponents.  In addition to the exponents defining
these power laws, another critical exponent, \viz\ the dynamic
exponent $z$, is required
for the singularities in the {\it dynamics}. This exponent $z$ is
defined by the relationship that holds between the correlation length
$\xi$ and the correlation time $\tau$, namely $\tau \propto \xi^z$.

One of the directions along which one can approach the critical
singularity is the finite-size direction, \ie, one increases the system
size $L$ while keeping the independent thermodynamic variables at their
infinite-system critical values. In this case, $\xi \propto L$ so
that $\tau \propto L^z$. This relation has been used extensively to
obtain the dynamic exponent $z$ from finite-size calculations.

In this paper we deal with universality of dynamic critical-point
behavior.  One would not expect systems in different static
universality classes to have the same dynamic exponents, and even
within the same static universality class, different dynamics may have
different exponents.  For instance, in the case of the Ising model,
Kawasaki dynamics which satisfies a local conservation law \cite{KD}
has a larger value of $z$ than Glauber dynamics,\cite{GD} in which
such a conservation law is absent.  Also the introduction of nonlocal
spin updates, as realized \eg\ in cluster algorithms, is known to lead
to a different dynamic universal behavior.\cite{SW,W,HB}

Conservation laws and nonlocal updates tend to have a large effect on
the numerical value of the dynamic exponents, but until fairly
recently, numerical resolution of the expected differences of dynamic
exponents of systems in different static universality classes for dynamics
with local updates has been elusive.  This is caused by the difficulty
of obtaining the required accuracy in estimates of the dynamic
critical exponent. Under these circumstances it is evident that only
a limited progress has been made with respect to the interesting
questions regarding dynamic universality classes.

In this paper we present a detailed exposition of a method of
computing dynamic exponents with high
accuracy.\cite{NBdyn.prl.96,NBdyn.prl.97} We consider single spin-flip
Glauber dynamics. This is defined by a Markov matrix, and computation
of the correlation time is viewed here as an eigenvalue problem, since
correlation times can be obtained from the subdominant eigenvalues
of the Markov matrix.  

If a thermodynamic system is perturbed out of equilibrium, different
thermodynamic quantities relax back at a different rates.  More
generally, there are infinitely many independent relaxation modes for
a system in the thermodynamic limit.  Let us label the models within
a given universality class by means of $\kappa$, and denote by
$\tau_{Li\kappa}$ the autocorrelation time of relaxation mode $i$
of a system of linear dimension $L$.  In this paper we
present strong numerical evidence that, as indeed renormalization
group theory suggests, at criticality the relaxation times have the
following factorization property
\begin{equation}
\tau_{Li\kappa} \,\approx\, m_{\kappa} A_i L^z,
\label{eq.factor.tau}
\end{equation}
where $m_{\kappa}$ is a {\it non-universal} metric factor, which
differs for different representatives of the same universality class
as indicated; $A_i$ is a {\it universal} amplitude which depends on
the mode $i$; and $z$ is the {\it universal} dynamical exponent
introduced above.

While the relaxation time of the slowest relaxation mode is obtained
from the second-largest eigenvalue of the Markov matrix, lower-lying
eigenvalues yield the relaxation times of faster modes.  To compute
these we construct, employing a \MC\ method, variational approximants
for several eigenvectors. These approximants are called optimized
trial vectors.  The corresponding eigenvalues can then be estimated by
evaluating with \MC\ techniques the overlap of these trial vectors and
the corresponding matrix elements of the Markov matrix in the truncated basis
spanned by these optimized trial vectors.  It
should be noted that both the optimization scheme and the evaluation
of these matrix elements critically depend on the fact that the Markov
matrix is sparse.  That is, the number of configurations accessible
from any given configuration is equal to the number of sites only,
rather than the number of possible spin configurations.

Given such fixed trial vectors, this approach has the advantage of
simplicity and high statistical accuracy, but the disadvantage is that
results are subject to systematic, variational errors, which only
vanish in the ideal limit where the variational vectors become exact
eigenvectors, or span an invariant subspace of the Markov matrix. Since the
condition is rarely satisfied in cases of practical interest,
a projection Monte Carlo method is then used, to reduce the
systematic error, but this is at the expense of an increase of the
statistical errors. The methods we use in this paper is a combination
and generalization of the work of Umrigar {\it et
  al.},\cite{CyrusOptimization} and that of Ceperley and Bernu.
\cite{CeperleyBernu88}

To summarize, the \MC\ method discussed here consists of two phases.
In the first phase, trial vectors are optimized.  The ultimate, yet
unattainable goal of this phase is to construct exact eigenvectors.
In this phase of the computation, very small \MC\ samples are used,
consisting typically of no more than a few thousand spin
configurations.  In the second phase, one performs a standard \MC\ 
computation in which one reduces statistical errors by increasing the
length of the computation rather than the quality of the variational
approximation.

The computed correlation times, derived from the partial solution of
the eigenvalue problem as sketched above, are used in a finite-size
analysis to compute the dynamic critical exponent $z$. We verify its
universality for several models in the static universality class of
the two-dimensional Ising model. We also address another manifestation
of dynamic universality. As was mentioned, in addition to the usual
static critical exponents, there is only one new exponent that governs
the leading singularities of critical dynamics, \viz., $z$.  Similarly,
one would expect that, within the context of Glauber
dynamics, the description of time-dependent critical-point amplitudes
requires only a single non-universal metric factor to determine the
time-scale of each different model within a given universality class.
Our results corroborate this idea, which is the immediate
generalization to critical dynamics of work on static critical
phenomena by Privman and Fisher.\cite{Privman.Fisher.84}

In this paper we apply the techniques outlined above to three
different two-dimensional Ising models subject to Glauber-like spin
dynamics.  These models are defined on a simple quadratic lattice of
size $L \times L$ with periodic boundary conditions. The Hamiltonian
$\H$, defined on a general spin configuration $S=(s_1,s_2,\dots)$, is
given by
\begin{equation}
\frac{\H(S)}{kT} = -K \sum_{\langle ij \rangle} s_i s_j 
            -K' \sum_{[kl]} s_k s_l,
\label{Ham}
\end{equation}
where the first summation is on all nearest-neighbor pairs of sites of
the square $L \times L$ lattice; the second summation is on all
next-nearest-neighbor pairs; the Ising variables $s_i,\dots,s_l$
assume values $\pm 1$.  Periodic boundaries are used throughout. In
particular, we focus on models described by three ratios $\beta=
K'/K$, namely $\beta=-1/4$, 0 (nearest-neighbor model) and 1
(equivalent-neighbor model). The non-planar models for $\beta \neq 0$
are not exactly solvable and their critical points are known only
approximately.  Yet, it was demonstrated to a high degree of numerical
accuracy that that they belong to the static Ising universality
class.\cite{BNuniv,NBPhys} For the nearest neighbor model the critical
coupling is $K={1 \over 2}\,\ln(1+\sqrt 2)$, for the other two models
estimates of the critical points are $K=0.1901926807(2)$ for $\beta=1$
and $K=0.6972207(2)$ for $\beta=-1/4$.\cite{NBPhys,NBdef}

We use the dynamics of the heat-bath algorithm with random site
selection.  The single-spin-flip dynamics is determined by the Markov
matrix $P$ defined as follows. The element $P(S',S)$ is the transition
probability of going from configuration $S$ to $S'$. If $S$ and $S'$
differ by more than one spin, $P(S',S)=0$. If both configurations
differ by precisely one spin,
\begin{equation}
P(S',S)= \frac{1}{2 L^2} \left\{ 
1 - \tanh \left[ \frac{ \H(S')-\H(S)}{2kT} \right] \right\},
\label{Mar}
\end{equation}
where $L^2$ is the total number of spins. The diagonal elements $P(S,S)$
follow from the conservation of probability
\begin{equation}
\sum_{S'} P(S',S)=1,
\label{cp}
\end{equation}
where $S'$ runs over all possible $2^{L^2}$ spin configurations. 

We denote the probability of finding spin configuration $S$ at time $t$
by $\rho_t(S)$.
By design, the stationary state of the Markov process is the
equilibrium distribution
\begin{equation}
\rho_{\infty}(S) =
\frac{\exp[-\H(S)/kT]}{Z} \equiv \frac{ \psi_{\rm B}(S)^2}{Z},
\end{equation}
where the normalization factor $Z$ is the partition function.

The dynamical process defined by Eq.~(\ref{Mar}) is constructed so as
to satisfy detailed balance, which is equivalent to the statement that
the matrix $\P$ with elements
\begin{equation}
\P(S',S) \equiv \frac{1}{\psi_{\rm B}(S')} P(S',S)\psi_{\rm B}(S)
\end{equation}
is symmetric. Therefore the eigenvalues of $P$ are real.

The Markov matrix determines the time evolution of $\rho_t(S)$,  \ie,
\begin{equation}
\rho_{t+1}(S)= \sum_{S'} P(S,S') \rho_t(S').
\label{eq.rhot}
\end{equation}
The simultaneous probability distribution $\rho_{t',t'+t}(S,S')$ that
the system is in state $S$ at time $t'$ and in state $S'$ at time $t'+t$
is
\begin{equation}
\rho_{t',t'+t}(S,S')=  P^t(S',S) \rho_{t'}(S),
\label{eq.rhott}
\end{equation}
where $P^t(S',S)$ denotes the $(S',S)$ element of the $t$-th power of
the matrix $P$. For sufficiently large times $t'$, one may take
$\rho_{t'}(S)= \rho_{\infty}(S)$ so that the autocorrelation function
$C_A(t)$ of an observable $A$, the average with respect to time $t'$ of
$A(t') A(t+t')$, can equivalently be written as the ensemble average
$\langle A(t') A(t'+t) \rangle$ for large $t'$.
Thus 
\begin{equation}
C_A(t) = \lim_{t'\to \infty}\sum_{S} \sum_{S'} A(S) A(S') \rho_{t',t'+t}(S,S')
\label{eq.autoc}
\end{equation}
where $A(S)$ denotes the value of $A$ in a spin configuration $S$.
After substitution of Eq.~(\ref{eq.autoc}) and expansion of
$A(S)\rho_{t'}(S)$ in right-hand 
eigenvectors of the Markov matrix, it follows at once that the
time-dependent correlation functions of a system of size $L$ have
the following form
\begin{equation}
C_A(t) =
\sum_i c_i\, \mbox{sgn}\,\lambda_{Li}^t\exp[-{t \over L^2\tau_{Li}}],
\label{eq.auto}
\end{equation}
where the dependence on the specific model $\kappa$ has been suppressed
in denoting by $\tau_{Li}$ the relaxation times of the independent
modes of the equilibration process. The $\tau_{Li}$ are determined by
the eigenvalues of the Markov matrix.  We denote these eigenvalues
$\lambda_{Li}$ $(i=0,1,2,\ldots,2^{L^2}-1)$, and order them so that
$1=\lambda_{L0}>|\lambda_{L1}|\geq|\lambda_{L2}|\geq\cdots $.  Note that
conservation of probability implies that $\lambda_{L0}=1$; by
construction, the corresponding right-hand eigenvector is the
Boltzmann distribution. 

The relaxation times are given by
\begin{equation}
\tau_{Li} = - \frac{1}{L^2 \ln |\lambda_{Li}|} 
\mbox{\hspace{10mm}} (i=1,2,\ldots)
\label{tau}
\end{equation}
The factor $L^2$ is inserted because, as usual, time is measured units
of one flip per spin, which corresponds to $L^2$ iterations of the
process described by Eq.~(\ref{eq.rhot}).

Note that the stochastic matrix ${P}$ has the same symmetry properties
as the Hamiltonian and the Boltzmann distribution. In addition to spin
inversion, these symmetries include translations, reflections and
rotations of the $L \times L$ lattice. It follows that each eigenvector
of ${P}$, as well as its associated relaxation mode, has distinct
symmetry properties that can be characterized by
a set of ``quantum numbers.'' For instance, the eigenvector associated
with the second-largest eigenvalue is antisymmetric under spin inversion,
and invariant under translations, reflections and rotations. It describes
the relaxation of the total magnetization; this process, with relaxation
time $\tau_{L1}$, is thus the slowest relaxation mode contained in the
stochastic matrix.

In this work, we restrict ourselves to relaxation modes that are invariant
under geometric symmetries of the spin lattice. However, in addition
to eigenvectors that are symmetric under spin inversion, we include
antisymmetric ones, so as to obtain the longest relaxation time.
As a consequence of this restriction to geometric invariance, 
spatially non-homogeneous relaxation processes fall outside the scope
of this work.

By design, the stationary state of the Markov process is the
equilibrium state
\begin{equation}
\rho_{\infty}(S) =
\frac{\exp[-\H(S)/kT]}{Z} \equiv \frac{ \psi_{\rm B}(S)^2}{Z},
\end{equation}
where the normalization factor $Z$ is the partition function.

The dynamical process defined by Eq.~(\ref{Mar}) is constructed so as
to satisfy detailed balance, which is equivalent to the statement that
the matrix $\P$ with elements
\begin{equation}
\P(S',S) \equiv \frac{1}{\psi_{\rm B}(S')} P(S',S)\psi_{\rm B}(S)
\end{equation}
is symmetric.

The layout of the rest of this paper is as follows.  In Section
\ref{sec.exhaust} we discuss the general principles of the method used
in this paper.  The expressions in this section feature exhaustive
summation over all spin configurations, which renders them useless for
practical computations.  The \MC\ summation methods employed instead
are discussed in Section \ref{sec.mc.summation}.  Trial vectors, a
vital ingredient of the method, are discussed in Section
\ref{sec.trial.states}, and numerical results are discussed in Section
\ref{sec.numerical}.  Finally, Section \ref{sec.discussion} contains a
discussion of issues that remain to be addressed by future work.

\section{EXACT SUMMATION}
\label{sec.exhaust}
\subsection{Variational approximation}
\subsubsection{A single eigenvector}
In this section we discuss trial vector optimization, which is used to
reduce the statistical errors in the \MC\ computation of eigenvalues
of the Markov matrix.  First, we review the case of a single
eigenvalue,\cite{CyrusOptimization,N96,NU96} and then we generalize to
optimization of multiple trial vectors.  In this section, we discuss
the exact expressions involving summation over all possible spin
configurations.  In cases of practical interest, these expressions
cannot be evaluated as written; for their approximate evaluation one
uses the Monte Carlo methods discussed in Section
\ref{sec.mc.summation}.

A powerful method of optimizing a single, many-parameter trial vector,
say $\psiTket$, is minimization of the variance of the {\it
  configurational eigenvalue,} which in the context of quantum Monte
Carlo is called the {\it local energy}.  That is, define
$\psiT(S)=\langle S\psiTket$ for an arbitrary configuration $S$.
We wish to satisfy the eigenvalue equation
\begin{equation} 
\psit {}'(S) = \lambda \psiT(S),
\label{eq.lamda1}
\end{equation} 
where the prime indicates matrix multiplication by $\P$, \ie, $f'(S)
\equiv \sum_{S'} \P(S,S') f(S')$ for any function $f$ defined on the
spin configurations. Even if $\psiT$ is not an eigenvector, one can define
the {\it configurational eigenvalue} by
\begin{equation}
\lambdac(S)=\left\{\begin{array}{l}
{\displaystyle \psit {}'(S) \over \displaystyle \psiT(S)}, \mbox{ if }
 \psiT(S)\ne 0\\ 0, \mbox{ otherwise.}
\end{array}\right. 
\label{eq.lambdac}
\end{equation}
If $\psiT$ is not an eigenvector, Eq.~(\ref{eq.lamda1}) gives an
overdetermined set of equations for $\lambda$, for a given $\psiT$ and
a sufficiently big set of configurations $S$.  One can obtain a
least-squares estimate of the eigenvalue $\lambda$ by minimizing the
squared residual of Eq.~(\ref{eq.lamda1}). This yields the
usual variational estimate
\begin{equation}
\lambda(p)={\psiTbra \P \psiTket \over \psiTbr \psiTket}=
{\sum_S \psit {} '(S) \psiT(S)
\over
\sum_S \psiT(S)^2}={\sum_S \lambdac(S) \psiT(S)^2
\over
\sum_S \psiT(S)^2},
\label{eq.lamdabar}
\end{equation}
which is the average of the configurational eigenvalue $\lambdac$.

The standard Rayleigh-Ritz variational method, which can be used for
the largest eigenvalue consists in maximization of $\blambda(p)$ with
respect to the parameters $p$.  However, one can formulate a different
optimization criterion as follows.  The gradient of $\blambda(p)$ with
respect to $\psiT(S)$ is
\begin{equation}
{\partial \blambda(p) \over \partial \psiT(S)}=
2{\psit {}'(S)-\blambda(p) \psiT(S) \over \sum_{S'} \psiT(S')^2}.
\label{eq.grad.lamda}
\end{equation}
Clearly, this gradient vanishes for {\em any} eigenvector and this
suggests as an alternative optimization criterion minimization of the
magnitude of the gradient of a normalized trial vector $\psiT$.  With
respect to Eq.~(\ref{eq.lamda1}), this corresponds to minimization of
the normalized squared residual
\begin{equation}
\chi^2(p) =
 {\sum_S [\psit {}'(S)-\blambda(p) \psiT(S)]^2 \over \sum_S \psiT(S)^2 }=
 {\sum_S [\lambdac(S)-\blambda(p)]^2 \psiT(S)^2 \over \sum_S \psiT(S)^2 }=
{\psiTbra (\P-\blambda)^2 \psiTket \over \psiTbr \psiTket}.
\label{eq.chi^2}
\end{equation}
which equals the variance of the configurational eigenvalue, as shown.

\subsection{Multiple eigenvectors}
\label{sec.more.states}

Minimization of $\chi^2(p)$ is a valid criterion for any eigenvector,
but if this is used without the equivalent of an orthogonalization
procedure, one would in practice simply keep reproducing an
approximation to the same eigenvector, the dominant one most of the time.
Since orthogonalization is not easily implemented with Monte Carlo
methods, we utilize straightforward generalizations of
Eqs.~(\ref{eq.lamda1}) and (\ref{eq.lamdabar}) to deal with more than
one eigenvalue and eigenvector.  Eq.~(\ref{eq.chi^2}) is a little
problematic in this respect, as will become clear.

Suppose we start from a set of $n$ trial vectors $\psit i$,
$(i=0,1,\cdots,n-1)$. We can then write Eq.~(\ref{eq.lamda1}) in
matrix form
\begin{equation}
\psit i'(S)=\sum_{j=0}^{n-1} \hLam_{ij} \psit j(S).
\label{eq.Lam_matrix}
\end{equation}
As before, the prime on the left-hand side indicates matrix
multiplication by $\P$, and again Eq.~(\ref{eq.Lam_matrix}) for all
$i$ and $S$ form an overdetermined set of equations for the matrix
$\hLam_{ij}$.  These equations have no solution, unless the $n$ basis
vectors $\psit i$ span an invariant subspace of the matrix $\P$, which
in non-trivial applications of course is never the case.  Again,
however, one can solve for the matrix elements $\hLam_{ij}$ in a
least-squares sense.  This yields
\begin{equation}
\hLam= {\hcP} \hN^{-1},
\label{eq.Lam=P/N}
\end{equation}
where
\begin{equation}
\hN_{ij}= \sum_S \psit i(S) \psit j(S)=
\psitbr i \psitket j,
\label{eq.N}
\end{equation}
and
\begin{equation}
{\hcP}_{ij}= \sum_S \psit i '(S) \psit j (S)=
\psitbra i \P \psitket j.
\label{eq.P}
\end{equation}
Note that although these matrix elements depend on the normalization
of the $\psitket i$, the matrix $\hLam$ is invariant under an overall
change of normalization.

By diagonalization of the $n \times n$ matrix $\hLam$ one obtains an
approximate, partial eigensystem of the Markov matrix.  More
specifically, suppose that
\begin{equation}
\hLam= D^{-1} \diag( \tilde\lambda_0,\dots,\tilde\lambda_{n-1}) D.
\end{equation}
The eigenvalues $\tlambda_0 > \tlambda_1 \geq \cdots \geq
\tlambda_{n-1}$ of $\hLam$ are variational lower bounds for the exact
eigenvalues of the Markov matrix $P$, in the sense that
$\tilde\lambda_i \le \lambda_i$,\cite{McD} if the exact eigenvalues
are numbered such that $\tlambda_0 > \tlambda_1 \geq \cdots$, in
contrast with the convention used in the discussion following
Eq.~(\ref{eq.auto}). This property is a consequence
of the interlacing property of the eigenvalues of symmetric matrices
and their submatrices, also known as the separation
theorem.\cite{interlacing} Note that in denoting the
eigenvalues we omit the index $L$ indicating system size, where this
is not confusing. The approximate eigenvectors $\tpsi_i$ are given by
\begin{equation}
\tpsi_i=\sum_{j=0}^{n-1} D_{ij} \psit j,
\label{eq.phi}
\end{equation}
which can be verified as follows: multiply Eq.~(\ref{eq.Lam_matrix})
through by $D_{ki}$, and sum on $i$ to verify that $\tpsi'_k$
proportional to $\tpsi_k$.
The expressions derived above are usually\cite{CeperleyBernu88}
derived by starting from the linear combinations given in this last
equation.  The $D_{ij}$ then are treated as variational parameters,
and are determined by requiring stationarity of the Rayleigh quotient.
This yields the following equation for the $D_{ij}$
\begin{equation}
\sum_j D_{ij} \hP_{jk} = \tlambda_i \sum_j D_{ij} \hN_{jk},
\label{eq.geneig}
\end{equation}
a generalized eigenvalue problem equivalent to the eigenvalue problem
defined by $\hLam$ defined in Eq.~(\ref{eq.Lam=P/N}).

Next we discuss the generalization to more than one trial vector of
minimization of the variance as given by Eq.~(\ref{eq.chi^2}).  In
this context it is important to keep in mind that the variational
approximation is invariant under replacement of the basis vectors by a
non-singular linear superposition.  This yields a similarity
transformation of $\hLam$, and leaves invariant the approximate
eigenvalues and eigenvectors.  Of course, one would like to have an
optimization criterion that shares this invariance.  The
squared residual of Eq.~(\ref{eq.Lam_matrix}) fails in this respect.
This sum is not even invariant under a simple rescaling of the basis
functions $\psit i$, and there is no obvious normalization comparable
to the one used in Eq.~(\ref{eq.chi^2}).

One way to perform the optimization in an invariant way is for each
choice of the optimization parameters to compute linear combinations
$\sum_j V_{ij} \psitket j$, where $V$ is the $n \times n$ 
matrix, such that $V\hLam V^{-1}$ is diagonal.  Each of these linear
combinations defines a $\chi^2$ via Eq.~(\ref{eq.chi^2}), and the
parameters in the basis functions can then be optimized by
minimization of these $n$ sums of squares.  One may define a convex
sum of these $\chi_i^2$ and optimize all parameters for all basis
functions simultaneously with respect to this combined object
function, or, as we did in our computations, one can perform the
optimization iteratively one vector at a time for eigenvalues with
increasing distance from the top of the spectrum.

Another approach that also yields invariant results is to perform
the optimization by dividing the set of configurations into several
subsets, and computing a matrix $\hLam$ for each subset.  One can then
minimize the variance of the eigenvalues over these subsets.  This is
the procedure we followed to produce the results reported in this
paper.

We have not investigated which of the two procedures described above
is superior.  Both do have a problem in common, namely that for a wide
class of variational basis vectors, they give rise to a singular or
nearly singular optimization problem.  This is a consequence of the
fact that the basis states are not unique, even if the eigenvalue
problem has a unique solution.  For the optimization problem this
means that there are many almost equivalent solutions, a problem
commonly encountered in (non-linear) when on performs least-squares
parameter fits.

More specifically, if the
basis vectors are such that a linear combination of trial vectors can
be expressed exactly (or to good approximation) in the same functional
form as the trial vectors themselves, then there is a gauge symmetry
(or an approximate gauge symmetry) that yields a class of equivalent
(or almost equivalent) solutions of the minimization problem.  That
is, if $\psitket i$ is a solution, $\sum_j V_{ij}\psitket j$ is an
equivalent (or almost equivalent) solution for any $V$.  This problem
can be solved straightforwardly by fixing the gauge and performing the
optimization subject to the constraint that $\sum_j V_{ij}\psitket
j\propto \psitket i$.  If the gauge symmetry holds only approximately,
this additional constraint may produce a sub-optimal solution.

\subsection{Beyond the variational approximation}

The eigenvalues obtained by the variational scheme discussed in the
previous sections have a bias caused by admixture of eigenvectors in
that part of the spectrum that is being ignored.  This variational
bias can be reduced in principle arbitrarily as
follows.\cite{CeperleyBernu88}

Let us introduce generalized matrices with elements
\begin{equation}
\hN_{ij}(t)=\psitbra i \P^t \psitket j
\label{eq.Nt}
\end{equation}
and
\begin{equation}
{\hcP}_{ij}(t)=\psitbra i \P^{t+1} \psitket j.
\label{eq.Pt}
\end{equation}

For $t=0$ these expressions reduce to Eqs.~(\ref{eq.N}) and
(\ref{eq.P}).  One can view the matrix elements for $t>0$ as having
been obtained by the substitution $\ket{\psit
  i}\to{\P}^{t/2}\ket{\psit i}$.  Expansion in the exact eigenvectors
immediately shows that the spectral weights are reduced of
``undesirable'' eigenvectors with less dominant eigenvalues, so that
the vectors ${\P}^{t/2}\ket{\psit i}$ span a more nearly invariant
subspace of $\P$ than the original states.  This process, however
becomes numerically unstable as $t\to \infty$, since in that
case all basis vectors of the same symmetry collapse onto the
corresponding dominant state.

\section{MONTE CARLO SUMMATION}
\label{sec.mc.summation}
Obviously, the summation over all spin configurations used in the
expressions in the previous section can, in general, be done only for
small systems. In this section, we discuss the Monte Carlo estimators
of the expressions presented above.  In principle, matrix multiplication
involves summation over all configurations and therefore is not practically
feasible.  However, for the dynamics we consider in this paper the
summation required for the matrix multiplication by $P$ in
$\langle S|\P\psiTket$ is an exception, since for a given $S$ there
are only $L^2$ configurations $S'$ from which $S$ can be reached with
one or fewer spin flips, and these are the only configurations for
which $P(S,S')$ does not vanish.  For all other configuration sums a
\MC\ method is used. 

To produce a \MC\ estimate of $\chi^2(p)$ as given in
Eq.~(\ref{eq.chi^2}), sample $\M$ spin configurations $S_\alpha$ with
$\alpha=1,\dots,\M$ from the Boltzmann distribution
$\psiB(S_\alpha)^2$. This yields a Monte Carlo estimate of $\blambda(p)$
\begin{equation}
\blambda(p) \approx{\sum_\alpha \hat\psit {}'(S_\alpha) \hat\psit {}
(S_\alpha) \over \sum_\alpha \hat\psit {}(S_\alpha)^2},
\label{eq.lamdabar.est}
\end{equation}
where
\begin{equation}
\hat\psit {}(S_\alpha)={\psiT(S_\alpha)\over \psiB(S_\alpha)},
\label{eq.psithat}
\end{equation}
\begin{equation}
\hat\psit {}'(S_\alpha)={\psit {}'(S_\alpha)\over \psiB(S_\alpha)}.
\label{eq.psithat1}
\end{equation}
Similarly,
\begin{equation}
\chi^2(p) \approx {\sum_\alpha
[\hat\psit {}'(S_\alpha)-\blambda\hat\psit {}(S_\alpha)]^2 \over
\sum_\alpha \hat\psiT(S_\alpha)^2}.
\label{eq.chi^2.est}
\end{equation}

Parameter optimization for a single vector is done by generating a
sample of a few thousand configurations and subsequently varying the
parameters $p$ while keeping this sample fixed.  The same applies to
the optimization of more than one vector, in which case estimates of
the required matrix elements $\hN_{ij}$ and $\hcP_{ij}$ are computed
by
\begin{equation}
\hN_{ij}\approx \sum_\alpha \hat\psit i(S_\alpha) \hat\psit j(S_\alpha)
\equiv \tN_{ij},
\label{eq.N.est}
\end{equation}
and
\begin{equation}
\hcP_{ij} \approx \sum_\alpha \hat\psit i '(S_\alpha) \hat\psit
j (S_\alpha)\equiv \tP_{ij}.
\label{eq.P.est}
\end{equation}

We attached tildes to the symbols on the right-hand side of
Eqs.~(\ref{eq.N.est}) and (\ref{eq.P.est}) to indicate that the
corresponding quantities are stochastic variables, which is important
to keep in mind for the following discussion.

Since the matrix $\hcP$ is symmetric, one might be inclined to
symmetrize its estimator $\tP_{ij}$ with respect to $i$ and $j$.  This
symmetrization, however, destroys the zero-variance principle
satisfied by the expressions as written.  As mentioned before, the
eigensystem of $\hLam$ is obtained exactly and without statistical
noise, if the basis vectors $\psit i$ are linear combinations of $n$
exact eigenvectors.  In that ideal case, $\hLam$ is uniquely
determined by Eq.~(\ref{eq.Lam_matrix}) even if it is applied only to
an subset of configurations $S$.  The same holds for a weighted subset
as represented by a \MC\ sample.  Even though the matrices $\hcP$ and
$\hN$ themselves depend on the weights and the subset, factors
responsible for statistical noise cancel in the product
$\hcP\hN^{-1}$.  To demonstrate this, we write the estimator in matrix
form
\begin{equation}
\tN = \tPsi \tPsi^{\dagger},
\end{equation}
and
\begin{equation}
\tP = \tPsi' \tPsi^{\dagger},
\end{equation}
where $\tPsi$ is a rectangular matrix with elements
$\tPsi_{i\alpha}=\hat\psit i (S_\alpha)$, and
$\tPsi_{i\alpha}'=\hat\psit i '(S_\alpha)$.  Eq.~(\ref{eq.Lam_matrix})
in matrix form becomes
\begin{equation}
\tPsi'=\hLam \tPsi.
\end{equation}
Clearly, if this last equation holds, $\tP \tN^{-1} = \tPsi'
\tPsi^{\dagger} ({\tPsi \tPsi^{\dagger}})^{-1}=\hLam$ without
statistical noise, as announced.

We have assumed that one matrix multiplication by the Markov matrix
can be done exactly; repeated multiplications rapidly become
intractable.  This is a problem for the computation of the matrix
elements given in Eqs.~(\ref{eq.Nt}) and (\ref{eq.Pt}).  To obtain a
statistical estimate of these matrix elements, one generates a time
series with the Markov matrix $P$.  One then exploits the fact that in
the steady state of the Markov process, the relative probability of
finding configurations $S_1,S_2,\dots,S_{t+1}$ in immediate succession
is given by
\begin{equation}
P(S_{t+1}|S_t)\cdots P(S_2|S_1) \psi_{\rm B}(S_1)^2.
\label{prdis}
\end{equation}

For a \MC\ run of length $\M$, this property allows us to write
\begin{eqnarray}
\hN_{ij}^{(t)}&=&
\sum_{S_0, \dots, S_t} \psit i (S_t) \hP(S_t|S_{t-1})\cdots
\hP(S_1|S_0) \psit j (S_0)\nonumber\\
&=&\sum_{S_0, \dots, S_t} \hat\psit i (S_t) \hat\psit j (S_0)
P(S_{t}|S_{t-1})\cdots P(S_1|S_0) \psiB(S_0)^2\\
&\approx&
\M^{-1} \sum_{\sigma=1}^{\M} \hat\psit i (S_{\sigma+t}) \hat\psit
j (S_{\sigma})
\label{eq.Nest}
\end{eqnarray}

Similarly,
\begin{eqnarray}
\hcP_{ij}^{(t)} &=&
\sum_{S_0, \dots, S_{t+1}} \psit i(S_{t+1})\, \hat P(S_{t+1}|S_{t})\cdots
\hat P(S_1|S_0) \psit j(S_0) \nonumber \\
&=& \sum_{S_0, \dots, S_t} 
\hat\psit i '(S_t) \hat\psit j (S_{0})\,
 \hat P(S_{t}|S_{t-1})\cdots \hat P(S_1|S_0) \psiB(S_0)^2\label{eq.Pij.xct}\\
&\approx& (2\M)^{-1} \sum_{\sigma=1}^{\M}
[\hat\psit i '(S_{\sigma+t}) \hat\psit j (S_{\sigma})+
 \hat\psit i '(S_{\sigma})\hat\psit j (S_{\sigma+t})].
\label{eq.Pij.mc}
\end{eqnarray}
The first term in expression~(\ref{eq.Pij.mc}) follows immediately
from expression~(\ref{eq.Pij.xct}); to obtain the second term one has
to use the time reversal symmetry of a stochastic process that
satisfies detailed balance, \viz,
\begin{equation}
P(S_{t+1}|S_t)\cdots P(S_2|S_1) \psi_{\rm B}(S_1)^2=
P(S_1|S_2)\cdots P(S_t|S_{t+1}) \psi_{\rm B}(S_{t+1})^2.
\label{eq.time_rvrs}
\end{equation}

Again, these estimators satisfy the zero-variance principle mentioned
above, as long as the expressions are used as written, \ie, without
symmetrization with respect to $i$ and $j$.

\section{TRIAL VECTORS}
\label{sec.trial.states}
As we mentioned, the form of the trial vectors used in these
calculations is a major factor determining the statistical accuracy of
the results.  It not too difficult to make an initial guess for the
form of the eigenvector corresponding to the second largest eigenvalue
of the Markov matrix.
Numerically exact calculations for small systems show that this
eigenvector is antisymmetric under spin inversion, which is a
manifestation of the longevity of fluctuations of the magnetization
and not a peculiarity of small systems.

\begin{figure}
\epsfig{figure=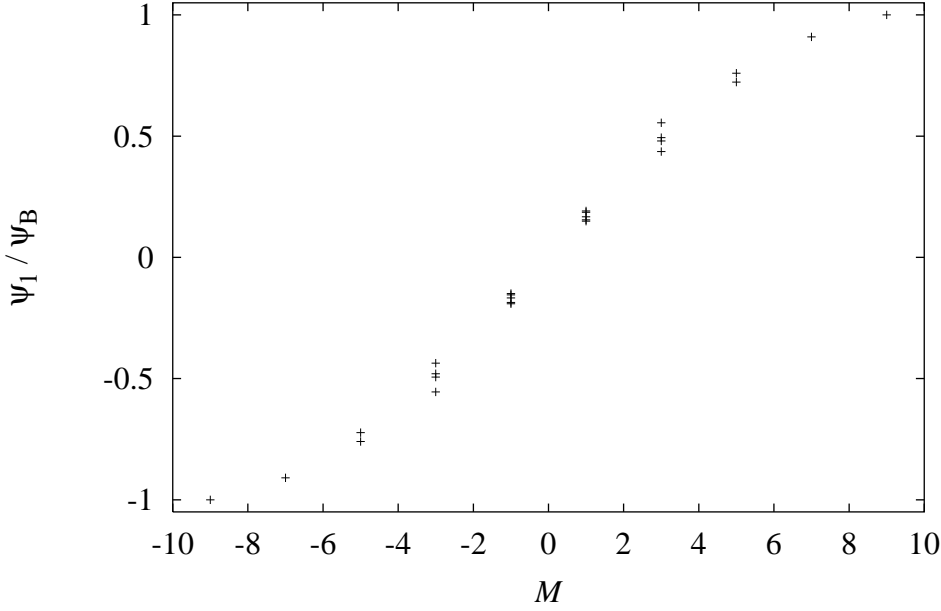}
\caption[xxx]
{Pre-factor $ \psi_1/ \psiB$ of the first sub-dominant eigenvector of
  the Markov matrix \vs\ total magnetization $M$ for a $3 \times 3$
  nearest-neighbor Ising ($\beta = 0$) lattice. For each value of $M$
  $\psi_{1}(S)/\psiB(S)$ is plotted for all configurations $S$ with
  $M(S)=M$.  All 512 points are plotted, but because of symmetries
  many coincide.}
\label{fig.psit13.vs.M}
\end{figure}

\begin{figure}
\epsfig{figure=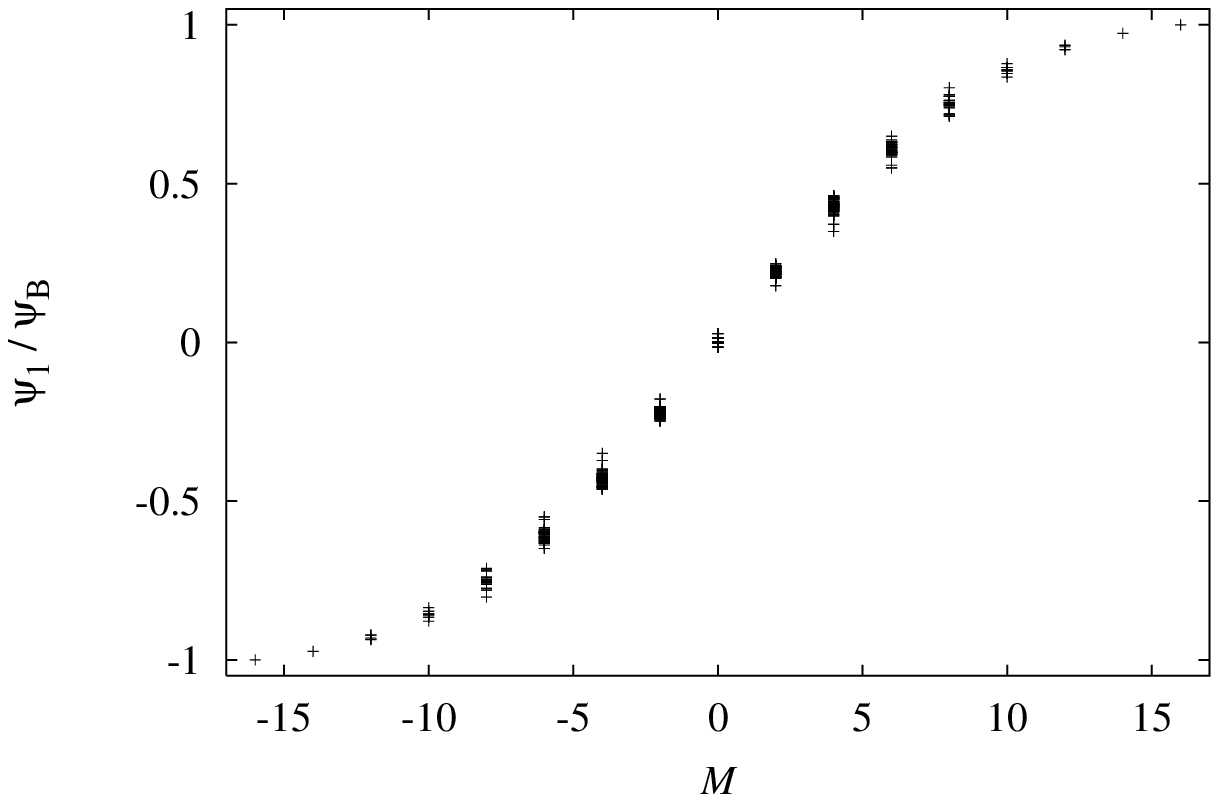}
\caption[xxx]
{Pre-factor $ \psi_1/ \psiB$ of the first sub-dominant eigenvector
  of the Markov matrix \vs\ total magnetization $M$ for a $4 \times 4$
  nearest-neighbor Ising ($\beta = 0$) lattice.}
\label{fig.psit14.vs.M}
\end{figure}

\begin{figure}
\epsfig{figure=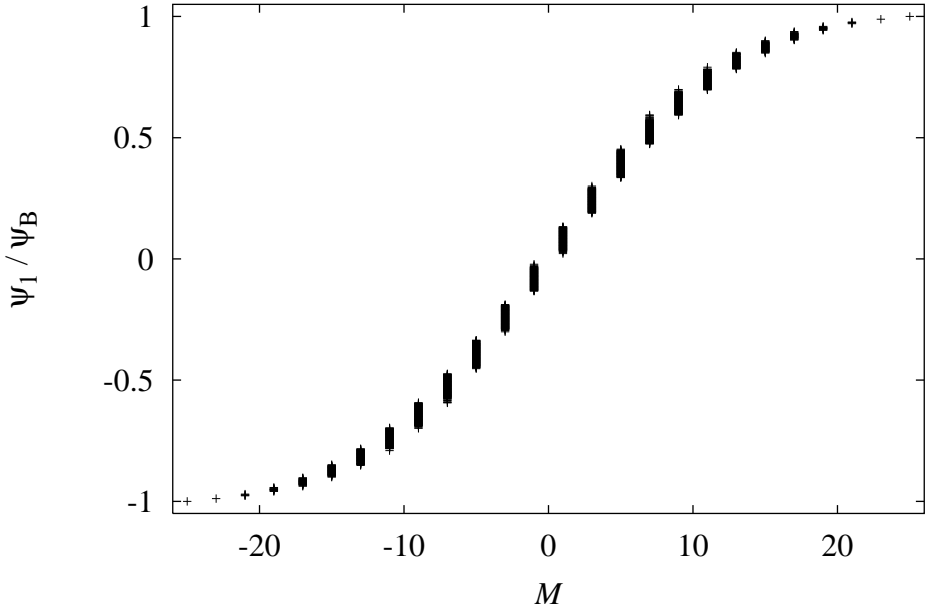}
\caption[xxx]
{Pre-factor $ \psi_1/ \psiB$ of the first sub-dominant eigenvector of
  the Markov matrix \vs\ total magnetization $M$ for a $5 \times 5$
  nearest-neighbor Ising ($\beta = 0$) lattice.  Although up to 400
  configuration collapse onto a single point, crowding prevents
  individual resolution of most data points.}
\label{fig.psit15.vs.M}
\end{figure}

This suggests the following initial approximation of the eigenvector
belonging to the second largest eigenvector, the first sub-dominant
eigenvector of the symmetrized Markov matrix $\P$:
\begin{equation}
\psit 1 (S)=m\,\psiB(S),
\label{eq.psit.of.m}
\end{equation}
where $m$ is the average magnetization.
Figs.~\ref{fig.psit13.vs.M}-\ref{fig.psit15.vs.M} are a plots of
$\psi_1(S)/\psiB(S)$ \vs\ the total magnetization $M=L^2 m$ for the
exact eigenvector $\psi_1$ computed for $3 \times 3$, $4 \times 4$,
and $5 \times 5$ nearest-neighbor Ising systems.  For all three, the
pre-factor $m$ in Eq.~(\ref{eq.psit.of.m}) clearly captures a
significant part of the truth, but there are two shortcomings.  First
of all, there is scatter, which indicates that $\psi_1(S)/\psiB(S)$ is
a function of more than just the magnetization.  Secondly, the
``curve'' is non-linear.  The latter problem can be cured quite easily
by replacing $m$ in Eq.~(\ref{eq.psit.of.m}) by an odd polynomial
$m(1+a_2 m^2+...)$ with coefficients $a_k$ to be determined
variationally.  

\begin{figure}
\epsfig{figure=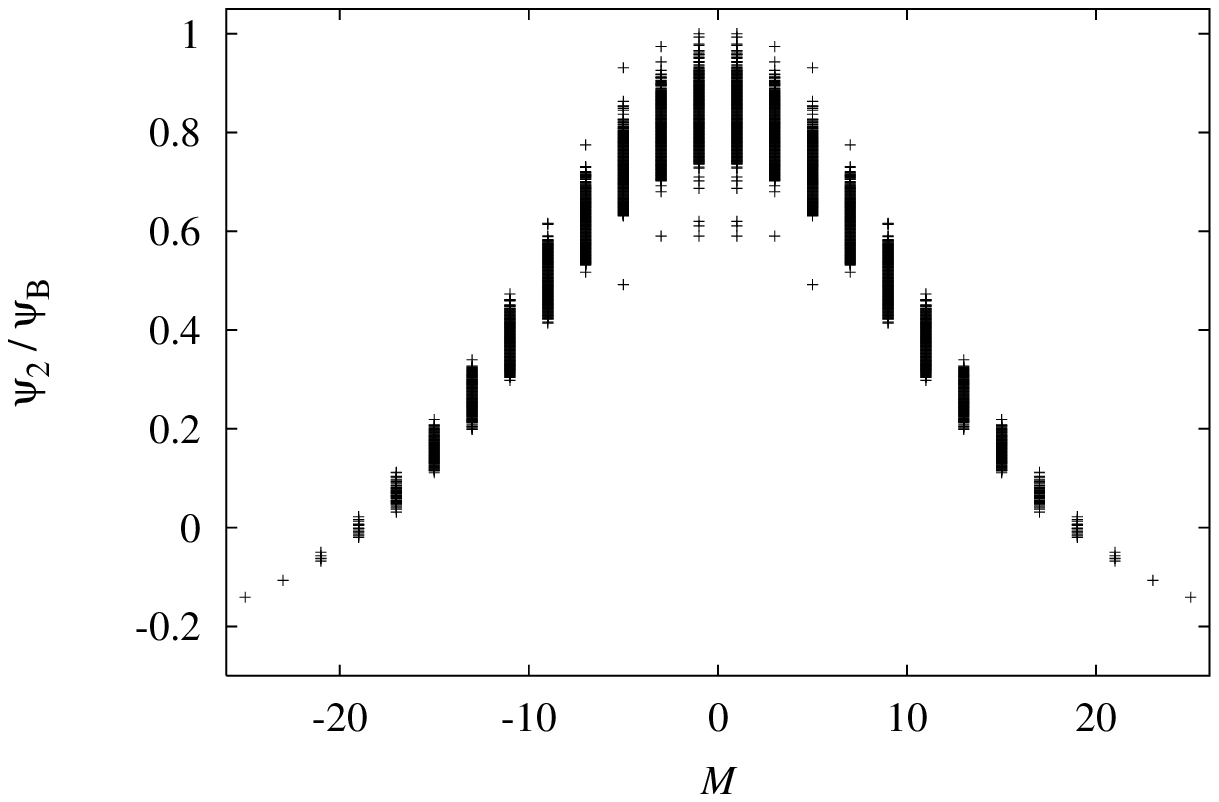}
\caption[xxx]
{Pre-factor $ \psi_2/ \psiB$ of the second sub-dominant eigenvector
  of the Markov matrix \vs\ total magnetization $M$ for a $5 \times 5$
  nearest-neighbor Ising ($\beta = 0$) lattice.}
\label{fig.psit25.vs.M}
\end{figure}

Similarly, computations for small systems (see Section~\ref{sec.exaxt}
for further details) suggest that the second largest eigenvalue is
associated with an eigenvector that is even under spin inversion, as
illustrated in Fig.~\ref{fig.psit25.vs.M}.  A trial vector of this
form is readily constructed by replacing $m$ on the right-hand side of
Eq.~(\ref{eq.psit.of.m}) by a polynomial even in $m$.  It turns out
that the general picture as just described is largely independent of
$L$.

\begin{figure}
\epsfig{figure=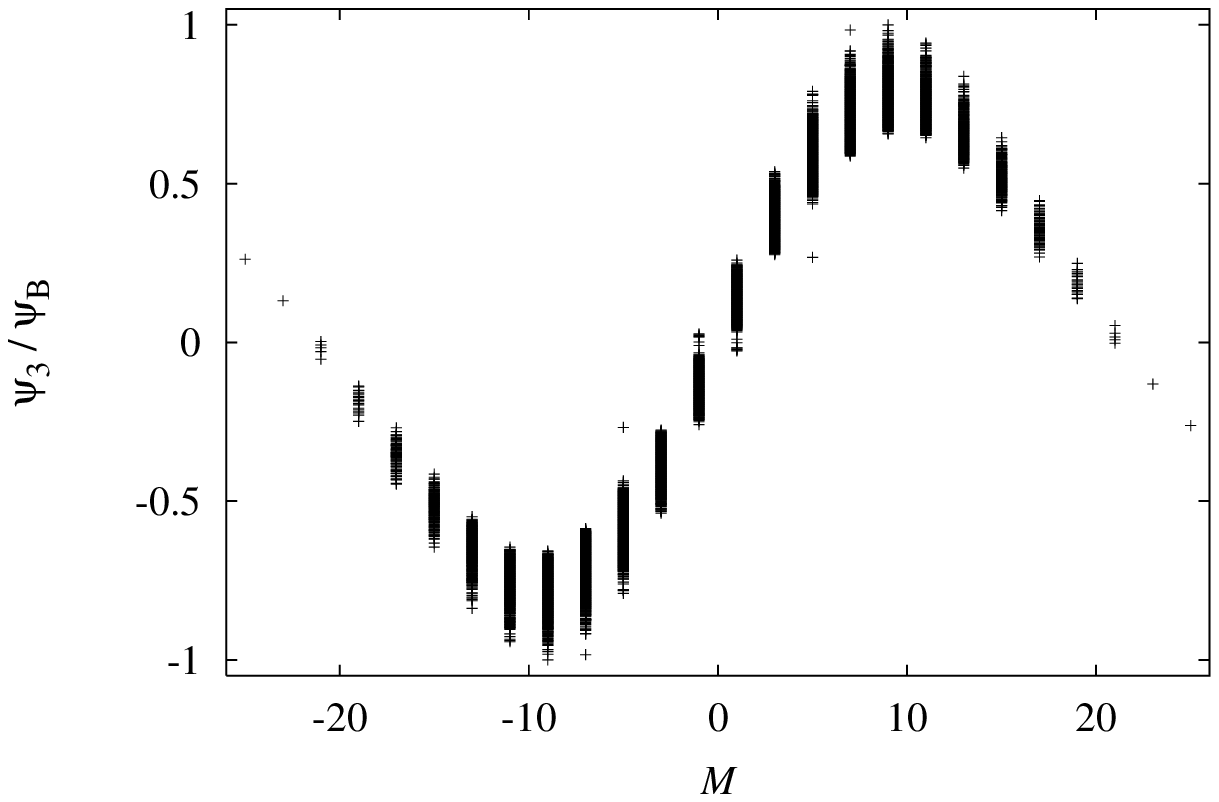}
\caption[xxx]
{Pre-factor $ \psi_3/ \psiB$ of the third sub-dominant eigenvector
  of the Markov matrix \vs\ total magnetization $M$ for a $5 \times 5$
  nearest-neighbor Ising ($\beta = 0$) lattice.}
\label{fig.psit35.vs.M}
\end{figure}

\begin{figure}
\epsfig{figure=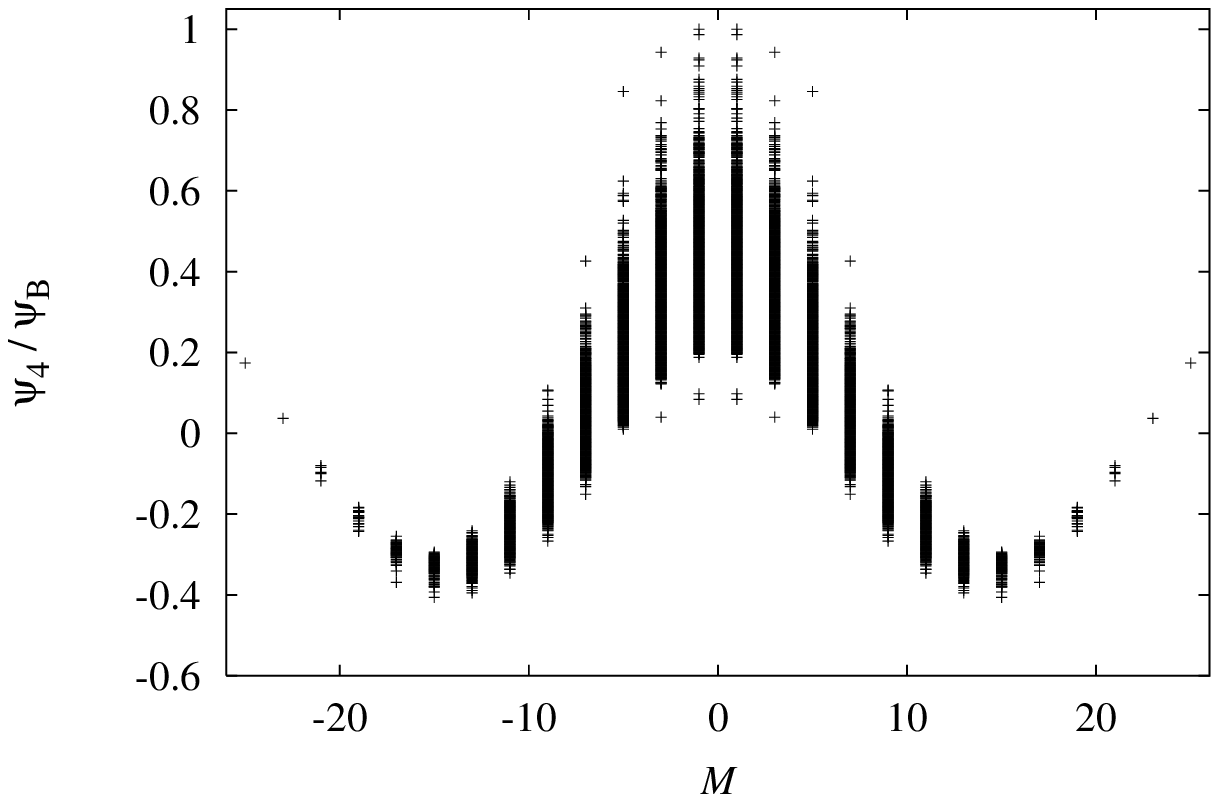}
\caption[xxx]
{Pre-factor $ \psi_4/ \psiB$ of the fourth sub-dominant eigenvector
  of the Markov matrix \vs\ total magnetization $M$ for a $5 \times 5$
  nearest-neighbor Ising ($\beta = 0$) lattice.}
\label{fig.psit45.vs.M}
\end{figure}

\begin{figure}
\epsfig{figure=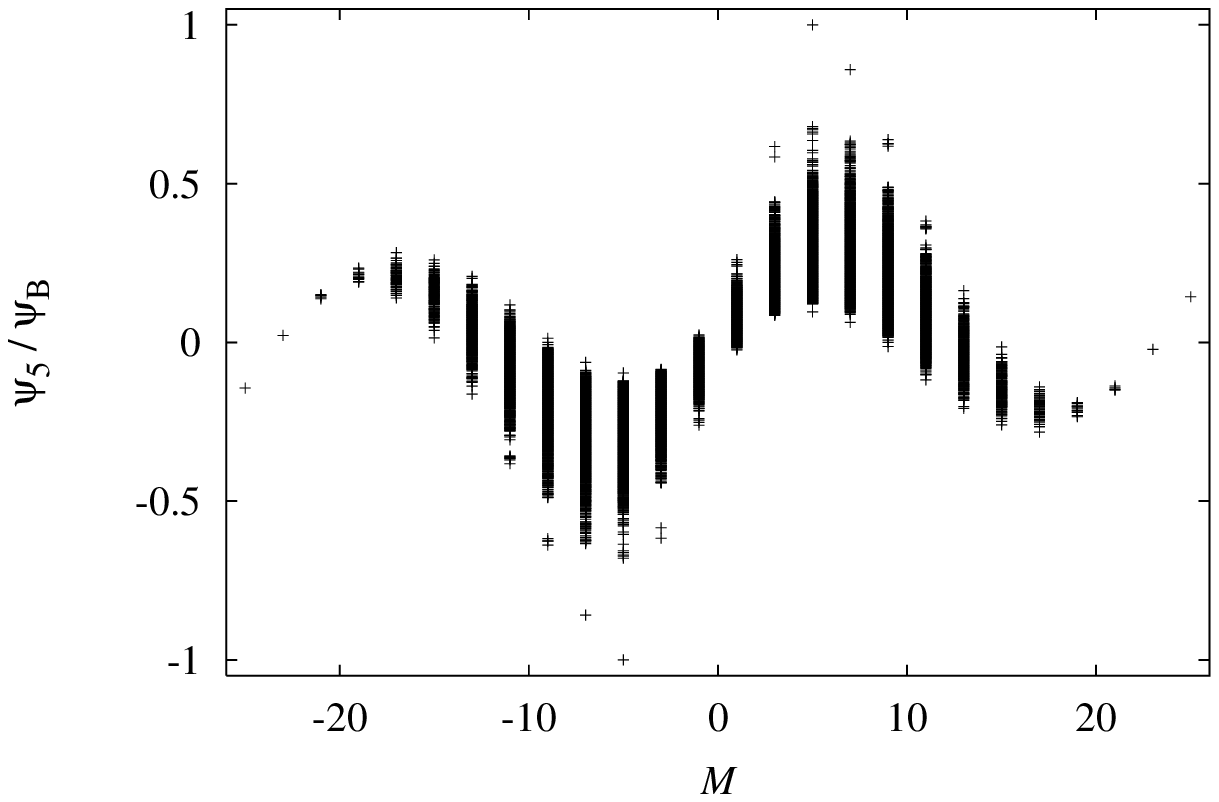}
\caption[xxx]
{Pre-factor $ \psi_5/ \psiB$ of the fifth sub-dominant eigenvector
  of the Markov matrix \vs\ total magnetization $M$ for a $5 \times 5$
  nearest-neighbor Ising ($\beta = 0$) lattice.}
\label{fig.psit55.vs.M}
\end{figure}

More in general, the plots shown in
Figs.~\ref{fig.psit13.vs.M}-\ref{fig.psit55.vs.M} strongly suggest
that the sub-dominant eigenvectors of the Markov matrix $P$, subject
to the imposed spin, rotation and translation symmetries, are
reasonably approximated by the Boltzmann distribution multiplied by a
mode-dependent function of the magnetization. As can be seen in
Figs.~\ref{fig.psit13.vs.M}-\ref{fig.psit55.vs.M}, the number of nodes
of this pre-factor increases by one as one steps down the spectrum,
but it is also clear that, especially for the less dominant
eigenvectors, the residual variance is significant.

To begin to address the problem of the scatter and to improve the
trial vector systematically, it is necessary to identify other
important variables besides the magnetization, and to incorporate them
in the trial vector.  We tried multi-spin correlations involving
nearby spins but after considerable failed experimentation we
established that long-wavelength fluctuations of the magnetization are
the suitable variables. This is reasonable when one compares \eg\ the
eigenvalue equations for $\psi_{0}(S')$ and $\psi_1(S')$ and realizes
that the eigenvalues differ only very little from unity except for
very small systems.  We therefore used the Fourier components of the
spin configuration, which are defined by
\begin{equation}
m_{\k}=L^{-2} \sum_{l_1=1}^L  \sum_{l_2=1}^L  \exp\left[{2 \pi i\over L}
(k_1 l_1+k_2 l_2)\right]\,s_{l_1l_2}
\end{equation}
where $\k=(k_1,k_2)$ with $0 \le k_1,k_2 < L$; and $s_{l_1l_2}$
denotes the spin at lattice site $(l_1,l_2)$.  Note that $m\equiv
m_{0,0}$.  If we restrict ourselves to eigenvectors that are
translationally invariant, the arguments presented in the previous
paragraph yield the following trial odd or even vectors:
\begin{eqnarray}
\psiT(S)&=& 
\psiB(S)\,\Big[a^{\pm}(m)+\mathop{{\sum}'}_{\k_1,\k_2} 
a^{\pm}_{\k_1 \k_2}(m)\,m_{\k_1} m_{\k_2} \,\delta_{\k_1+\k_2,0}
\nonumber\\
&+&\mathop{{\sum}'}_{\k_1,\k_2,\k_3} a^{\mp}_{\k_1 \k_2 \k_3}(m)
\,m_{\k_1} m_{\k_2} m_{\k_3} \,\delta_{\k_1+\k_2+\k_3,0}+\dots \Big]
\label{eq.psit}
\end{eqnarray}
The primes attached to the summation signs indicate that terms with
$\k_i = (0,0)$ are excluded.  The coefficients $a^{\pm},\ 
a^{\pm}_{\k_1 \k_2},\dots$ are polynomials in $m$, which are either
odd or even under spin inversion and are to be chosen according to the
desired symmetry.  Rotation and reflection symmetries
of the lattice are imposed by equating coefficients of the appropriate
monomials in $m$.

The results reported in Ref.~\onlinecite{NBdyn.prl.96} were obtained
using a more complicated version of Eq.~(\ref{eq.psit}), namely
\begin{eqnarray}
\psiT(S)&=& 
\psiB(S)\,\Big[a^-(m)+\mathop{{\sum}'}_{\k_1,\k_2} 
a^-_{\k_1 \k_2}(m)\,m_{\k_1} m_{\k_2} \,\delta_{\k_1+\k_2,0}
\nonumber\\
&+&\mathop{{\sum}'}_{\k_1,\k_2,\k_3} a^+_{\k_1 \k_2 \k_3}(m)\,m_{\k_1}
m_{\k_2} m_{\k_3} \,\delta_{\k_1+\k_2+\k_3,0}+\dots \Big]\nonumber\\
&\times&\Big[a^+(m)+\mathop{{\sum}'}_{\k_1,\k_2} 
a^+_{\k_1 \k_2}(m)\,m_{\k_1} m_{\k_2} \,\delta_{\k_1+\k_2,0}
\nonumber\\
&+&\mathop{{\sum}'}_{\k_1,\k_2,\k_3} a^-_{\k_1 \k_2 \k_3}(m)\,
m_{\k_1} m_{\k_2} m_{\k_3} \,\delta_{\k_1+\k_2+\k_3,0}+\dots \Big]
\label{eq.psit2}
\end{eqnarray}
In those calculations also the coupling constant appearing in the
Boltzmann factor was treated as variational parameter, but it turned
out that the optimal value of this parameter was indistinguishable
from the critical coupling.  It does not seem that the more
complicated form of expression (\ref{eq.psit2}) resulted in a major
improvement, but we did not perform a systematic comparison of these
trial vectors.

The coefficients in the trial vector are treated as variational
parameters.  As in all non-linear fitting problems it is important to
use parameters parsimoniously, and to do so one has to establish a
hierarchy among these parameters.  The scheme we used was to iterate
the following step: {\it (a)} systematically add terms of increasing
degree in $m$; {\it (b)} when this saturates increase the degree of
terms with products of $m_{\k}$ with $m_{\k} \ne (0,0)$.

The effectivity of this variational approach using low-momentum
Fourier components, as described here, becomes apparent when one
compares the variational eigenvalues with the exact numerical ones.
For instance, the difference in the case of the second eigenvalue of
the $L=5$ nearest-neighbor model was only $2\times 10^{-7}$.

\section{NUMERICAL RESULTS}
\label{sec.numerical}
\subsection{Exact eigenvectors for small systems}
\label{sec.exaxt} 
The full, symmetric Markov matrix $\hP$ for an $L \times L$ Ising
model is a $2^{L^2} \times 2^{L^2}$ matrix, so that exact numerical
calculations are possible only for very small systems; see \eg\ 
results for $L\leq 4$ in Ref.~\onlinecite{NBPhysica.80}. In the
present work, we performed such exact computations for systems up to
$L=5$.  In order to restrict the numerical task, we chose
representations of $\P$ in subspaces with the appropriate symmetries.
Two distinct symmetries were chosen, both of which impose invariance
of the eigenvectors of $\P$ with respect to geometric translation,
rotation and mirror inversion.  The vectors were chosen to be either
even or odd under spin inversion.  This reduced by almost a factor 400
the dimensionality of $\P$.  In this way, the computation of a
restricted set of eigenvectors became feasible for the resulting
matrices of order 86,056 for the $L=5$ cases. For the diagonalization
we made use of sparse-matrix methods and the conjugate-gradient method
(see e.g., Ref.~\onlinecite{NiKNW78,NVM93}) which computes the
eigenvector with the largest eigenvalue.  Subsequent orthogonalization
with respect to this eigenvector yields the eigenvector with the
second largest eigenvalue, and further eigenvectors can be obtained
similarly. Thus we obtained exact numerical solutions for six
eigenvalues $\lambda_{Li}$ and their corresponding eigenvectors
$\psi_{i}(S)$ ($i=0,\cdots,5$) of the eigenvalue equation
\begin{equation}
\sum_{S'} \P(S,S') \psi_{i}(S') = \lambda_{Li} \psi_{i}(S).
\label{eq.evprob}
\end{equation}
The largest eigenvalue $\lambda_{L0}$ is equal to 1, in accordance
with the conservation of probability; its corresponding eigenvector
satisfies $\psi_{0}(S)=\psiB(S)$, as follows from detailed balance. It
is even with respect to spin inversion: $\psi_{0}(S)= \psi_{0}(-S)$
where $-S$ is obtained from $S$ by inverting all spins. For all system
sizes and models included here, we observed that the six leading
eigenvectors, ordered according to magnitude of their eigenvalues,
alternate between the odd and even subspaces: the first eigenvector is
even, the second one is odd, the third one is even subspace, and so
on, with the caveat that for $L=2$ \eg\ the odd subspace contains only
two independent states.  As we discussed above, the resulting
eigenvectors provide useful information on how to construct trial
vectors; moreover, the knowledge of accurate eigenvalues for $L \leq
5$ provided an powerful test of the Monte Carlo method, the results of
which are presented in the following subsection.

\subsection{Monte Carlo calculations}
All simulations took place at the respective critical points of the
models considered.  This point is known exactly in the case of the
nearest-neighbor model [$K_c= \ln(1+\sqrt 2)/2$], and was determined
numerically \cite{NBPhys} for the other two models:
$K_c=0.1901926807\,(2)$ for the equivalent-neighbor model, and
$K_c=0.6972207\,(2)$ for the model with antiferromagnetic next-nearest
neighbor interactions. The finite-size scaling analysis presented in
Ref.~\onlinecite{NBPhys} showed that, to the extent they are
compatible with the numerical results, deviations from Ising universal
behavior are extremely small.  The raw simulation data used in this
current paper include the data on which were based the numerical
results for the largest relaxation time of the nearest-neighbor model,
reported in Ref.~\onlinecite{NBdyn.prl.96}. The latter results were
obtained from $8 \times 10^8$ Monte Carlo samples for systems with
finite sizes up to $L=15$.  The trial vector used for these
computations consisted was of the form given in expression
(\ref{eq.psit2}) and used up to 36 variational parameters.
Also included in the present analysis are the simulations reported in
Ref.~\onlinecite{NBdyn.prl.97}, which contained $1.2 \times 10^8$
Monte Carlo samples for all three models with system sizes up to
$L=20$.

In addition, new simulations for each of the three models were
performed, with a length of $2 \times 10^8$ Monte Carlo samples for
system sizes up to $L=20$, and of $1.6 \times 10^8$ Monte Carlo
samples for system size $L=21$. These new simulations used up to 89
variational parameters in the trial functions for each eigenvector of
each model. 

In order to suppress biases due to deviations of randomness, we made use
of a random number generator which combines two different binary shift
registers such as described and discussed in Ref. \onlinecite{BLH}.

The required Fourier components of the spatial
magnetization distribution were sampled at intervals of one sweep for
the smallest systems up to about 15 sweeps for the largest ones.  The
Monte Carlo calculation of the autocorrelation times (actually the
eigenvalues of the Markov matrix) was performed for each run as a
whole as well as separately for a number of up to 1024 blocks into
which the run was split. This blocking procedure enabled us to
estimate the statistical errors.  Furthermore, the calculation of the
eigenvalues according to $\hcP(t) \hN(t)^{-1} = \hLam^{(t)}$ still
depends on the time displacements $t$ [see Eqs.~(\ref{eq.Nest}) and
(\ref{eq.Pij.mc})]. The calculation of $\hLam^{(t)}$ was performed for
time displacements $tL^{-2}=0,1,2,\cdots$ up to 10 or 20 of the
above-mentioned intervals.  For small $t$ these eigenvalue estimates
reflect variational bias due to the residual contributions of
relaxation modes decaying faster than the mode for which the trial
vector was constructed. If the relaxation times of these faster modes
are considerably shorter than that of the mode under investigation,
one can clearly see a fast convergence of the eigenvalue estimate as a
function of $t$.  Convergence, however, occurs to a level that is only
approximately constant because of the correlated statistical noise
whose effect still depends on $t$. With increasing $t$, one can also
observe that the statistical errors increase.  The latter effect,
which is as slow as the pertinent relaxation mode, occurs when the
autocorrelations of the Monte Carlo sample are decreasing
significantly with $t$.  This situation was indeed observed for the
largest eigenvalues; the data converged well with $t$ before the
coherence of the sampled data was lost. It was thus rather simple to
select a ``best estimate'' of those eigenvalues. However, the
situation for the smaller eigenvalues investigated here was much more
difficult, because the relative differences between subsequent
autocorrelation times are much smaller.  The numerical results for the
eigenvalues are listed in Appendix A.

\subsection{Determination of the dynamic exponent}

In two-dimensional Ising models, finite-size corrections are known that
decay with finite size as $L^{-2}$, and integral powers thereof may also
be expected. In the absence of information on possible additional 
finite-size corrections of a different type that could occur in dynamic
phenomena, we try to describe the finite-size data for the various
autocorrelation times, as given in Eq.~(\ref{tau}),
by the formula
\begin{equation}
\tau_L \approx L^z \sum_{k=0}^{n_{\rm c}} \alpha_k L^{-2k}.
\label{fit}
\end{equation}
Here $z$ is the dynamic exponent, $\alpha_k$ the finite-size
amplitude, and $n_c$ is the number of correction-to-scaling terms
included.  Not explicitly shown in this notation is that the
autocorrelation times depend on the relaxation mode and
the model.

On the basis of Eq.~(\ref{fit}), a considerable number of
least-squares fits were applied to the numerical results for the
autocorrelation times. For each model and relaxation mode, one may
vary both $n_{\rm c}$, the number of correction terms, and the low-$L$
cutoff specifying the minimal system size included in the fit. The
smaller the number of corrections, the larger the low-$L$ cutoff must
be chosen in order to obtain an acceptable squared residual $\chi^2$.
A selection of fits that display the numerical trends is presented in
Appendix B.

The ``best fits'' were chosen on the basis of the $\chi^2$ criterion,
the dependence on the low-$L$ cutoff, and the mutual consistency of
fits with different $n_{\rm c}$. The fits are summarized in Table
\ref{tab:zbest}. Since the errors are not only of a statistical
nature, but also depend on residual bias in the autocorrelation times
and subjective choices made in the selection of the best fits, we
quote error bars equal to two standard deviations as obtained from
statistical considerations only. We believe that these two-sigma error
estimates are conservative in the case of the analysis of the second
and third largest eigenvalues of the $\beta \geq 0$ models.  The
$\beta=-1/4$ model was found to be numerically less well-behaved: the
statistical errors, as well as the corrections to scaling appear to be
larger. Also, the construction of trial vectors was somewhat less
successful than in the cases of the $\beta \geq 0$ models.

\begin{table}[t]
\caption{Best estimates for the dynamic exponent $z$ for five
relaxation modes in three Ising-like models. These results were
selected from a much larger set of least-squares fits, obtained 
for different choices of the minimum system size and of the 
number of corrections taken into account (see Appendix B). The
error estimate in the last decimal place of each entry is listed 
in parentheses, and is taken to be two standard deviations
in the best fit.
}
\vskip 1 ex
\begin{center}
\begin{tabular}{||l|lr|lr|lr||}
mode &$\beta=-1/4$&     &$\beta=0$&      &$\beta=1$&     \\
\hline
  1  & 2.164      & (3) & 2.1660  & (10) & 2.1667  & (5) \\
  2  & 2.166      & (3) & 2.167   &  (1) & 2.167   & (1) \\
  3  & 2.164      & (5) & 2.170   &  (2) & 2.167   & (1) \\
  4  & 2.17       & (1) & 2.162   &  (4) & 2.170   & (8) \\
  5  & 2.15       & (2) & 2.17    &  (1) & 2.17    & (1) \\
\end{tabular}
\end{center}
\label{tab:zbest}
\end{table}

The new data are somewhat more accurate than, and consistent with our
previous work.\cite{NBdyn.prl.97} They are also consistent with the
results of Wang and Hu \cite{Hu} for the slowest relaxation mode of a
different set of Ising-like models. They provide a clear confirmation
of universality of the dynamic exponent, with regard to relaxation
modes as well as models. Our best estimate $z=2.1667\,(5)$ applies to
the slowest (odd) relaxation mode of the equivalent-neighbor
($\beta=1$) model.

This result for $z$ is consistent with most of the recently published
values. This agreement includes the results of Stauffer \cite{Stauf1}
on damage spreading in the Ising model. (The value listed in
Ref.~\onlinecite{NBdyn.prl.96} was incorrectly quoted, for which we
apologize.) It is slightly larger than the value 2.14 derived by
Alexandrowicz~\cite{Awicz} on the basis of a scaling argument.

Next we address the question whether the finite-size divergence of the
autocorrelation times at criticality can be described by a dynamic
exponent $z=2$ when a logarithmic factor is included. This possibility
was suggested by Domany,\cite{Dom} and pursued by Swendsen \cite{Swe}
and Stauffer\cite{Stauf3}, who used very large lattices and found that
this possibility seems inconsistent with one way of simulation, but not
with a different one. Further references concerning this question are
given in Ref.~\onlinecite{Stauf2}. 

Although the present work is restricted to very small system sizes,
the data are relatively accurate.  Thus we tried to fit the following
form to the finite-size data for the slowest relaxation mode
\begin{equation}
\tau_L \approx L^z (1+b \ln L) (\sum_{k=0}^{n_{\rm c}} \alpha_k L^{-2k}).
\label{lfit}
\end{equation}
Fixing $z=2$ and taking $b$ as a variable parameter, we found that this
form could well describe the data for large enough $n_{\rm c}$. The 
quality, as determined by the $\chi^2$ criterion, of a number of 
such fits is shown in Table \ref{tab:logfit}. For comparison we 
include fits in which $z$ is a variable parameter without a logarithmic
correction.

\begin{table}[htbp]
\caption{
Comparison between fits to the autocorrelation times, with and without
a logarithmic finite-size dependence. These fits apply to the slowest
relaxation mode. The first column shows the minimum system size
included, the second the number of correction terms included.  The
fourth column displays the squared residual $\chi^2_z$ obtained when the
dynamic exponent $z$ was left free and the amplitude $b$ of the
logarithm was fixed at zero.  The fifth column shows the squared residual
$\chi^2_b$ when $z$ was fixed at value 2, while $b$ was left free. The
sixth column lists the number of degrees of freedom of the fit for
comparison.}

\vskip 1 ex
\begin{center}
\begin{tabular}{||c|c|c|c|c|c||}
$L\geq$ & model   & $n_{\rm c}$ & $\chi^2_z$ & $\chi^2_b$ & $d_f$\\
\hline
    5   &$\beta=0$&  1          &   239.     &    274.    & 76 \\
    6   &$\beta=0$&  1          &    98.     &    187.    & 72 \\
    7   &$\beta=0$&  1          &    83.5    &    127.    & 67 \\
    8   &$\beta=0$&  1          &    68.7    &     87.9   & 62 \\
    9   &$\beta=0$&  1          &    63.8    &     71.9   & 57 \\
   10   &$\beta=0$&  1          &    55.9    &     57.5   & 52 \\
\tableline
    8   &$\beta=1$&  1          &    56.0    &     71.1   & 51 \\
    9   &$\beta=1$&  1          &    50.1    &     57.4   & 47 \\
   10   &$\beta=1$&  1          &    49.4    &     49.7   & 43 \\
\tableline
    4   &$\beta=0$&  2          &    89.2    &    341.    & 76 \\
    5   &$\beta=0$&  2          &    85.0    &    136.    & 75 \\
    6   &$\beta=0$&  2          &    83.3    &     97.2   & 71 \\
\end{tabular}
\end{center}
\label{tab:logfit}
\end{table}

The results in Table \ref{tab:logfit} indicate that the fits with a
variable exponent are usually better than those which include a
logarithmic term, \ie, the residual $\chi^2$ decreases faster when the low-$L$
cutoff is increased. This is especially apparent for small $n_{\rm c}$
and for the $\beta=0$ and $\beta=1$ models, where the statistical
accuracy is optimal.

Finally we tried a fit according to Eq. (\ref{lfit}) with both $z$ 
and $b$ as free parameters. The resolution of both parameters
simultaneously is quite hard, and lies near the limit of what
can be gleaned from the present data. For the nearest-neighbor model
we find, using system sizes $L=8$ and larger, and $n_{\rm c}=1$,
that $z=2.13 \pm 0.07$ and $b=0.05 \pm 0.09$, which again fails to
support the presence of a logarithmic term.

\subsection{Universality of finite-size amplitudes}

In order to determine the leading amplitudes $\alpha_0$ more
precisely, we repeated the fits as used for the determination of the
dynamic exponent, but with the value of the latter fixed at $z=13/6$.
We note that the combined results for $z$ are consistent with this
fraction.  A considerable number of fits were made, and ``best
estimates'' of the amplitudes are presented in Table \ref{tab:abest}.

\begin{table}[t]
\caption{ Best estimates for the finite-size amplitudes of five
relaxation modes in three Ising-like models. These results were
selected from a much larger set of least-squares fits obtained 
for different choices of the minimum system size and of the 
number of corrections taken into account (see Appendix B). The
error estimate in the last decimal place of each entry is listed 
in parentheses, and is taken to be two standard deviations
of the best fit. The amplitudes can be written as the product of 
mode-dependent and model-dependent constants (see text); the
difference in the last decimal place between the amplitudes and 
this product is shown between square brackets.
}
\vskip 1 ex
\begin{center}
\begin{tabular}{||l|lrr|lrr|lrr||}
mode/model&$\kappa=1$&    &    &$\kappa=2$&    &   &$\kappa=3$&   &   \\
\hline
1 (odd)   & 6.763   & (6) &[-10]& 4.4089  &(13)& [8]& 2.8312  &(5)&[0]\\
2 (even)  & 0.2516  & (2) &  [2]& 0.16364 & (5)& [0]& 0.10510 &(2)&[0]\\
3 (odd)   & 0.1188  & (1) &  [1]& 0.07727 & (3)& [0]& 0.04963 &(1)&[0]\\
4 (even)  & 0.07195 & (7) &  [3]& 0.04677 & (4)&[-4]& 0.03008 &(3)&[2]\\
5 (odd)   & 0.0466  & (1) & [-1]& 0.03041 & (3)&[-1]& 0.01956 &(2)&[2]\\
\end{tabular}
\end{center}
\label{tab:abest}
\end{table}

As mentioned in Ref.~\onlinecite{NBdyn.prl.97}, according to a modest
generalization of accepted ideas on universality, the finite-size
amplitudes of the autocorrelation times should satisfy $\alpha_0=A_i
m_{\kappa}$, where the $m_{\kappa}$ are non-universal, model-dependent
constants; the subscript $\kappa$ refers to the specific model. We use
the notation $\kappa=2+{\rm sgn}(\beta)$ so that $\kappa=1$ refers to
the model with ferromagnetic nearest- and antiferromagnetic next
nearest neighbor couplings, $\kappa=2$ denotes the nearest-neighbor
model, and $\kappa=3$ to the equivalent-neighbor model.
The $A_i$, $i=1,\cdots,5$ are mode-dependent constants, whose ratios
are universal. Since only the product matters, we are free to choose
an arbitrary value for one of these constants. We chose to fix $A_1=1$,
so that all $A_i$ should become universal constants.

The remaining constants $A_i$ ($i=2,\cdots,5$) and $m_{\kappa}$ were
fitted accordingly to the amplitudes listed in Table \ref{tab:abest}.
The result of this least-squares fit is
$m_1 =  6.773     \pm  0.003$,
$m_2 =  4.4081    \pm  0.0009$,
$m_3 =  2.8312    \pm  0.0005$,
$A_2 =  0.037123  \pm  0.000008$,
$A_3 =  0.017530  \pm  0.000004$,
$A_4 =  0.010618  \pm  0.000006$, and
$A_5 =  0.006901  \pm  0.000005$.
This fit has 8 degrees of freedom and $\chi^2=9.0$, in a good agreement
with the assumptions of dynamic universality, and suggesting that 
our two-sigma error estimates are not unrealistic. The differences
between our amplitude estimates and the fitted product $A_i m_{\kappa}$
are included in Table \ref{tab:abest}, in units of the last decimal
place listed.


\section{DISCUSSION}
\label{sec.discussion}
The analysis presented above produces an apparently highly accurate
estimate of the dynamic critical exponent $z=2.1667 \pm 0.0005$ for
the case of the equivalent neighbor model, where the statistical
accuracy and the convergence are best. However, even in this case it
is not possible to rule out a divergence of the relaxation time of the
form $L^2(1+b \ln L)$.  Nevertheless, our results viewed in their
totality make this behavior rather unlikely.  Firstly, the assumption
of this logarithmic form yields fits that converge less rapidly, as
mentioned above.  Secondly, one would have to have $b \approx 1/6$
universally, independent of model and relaxation mode, since our
results for the range of system size we studied are consistent with a
divergence of the form $\tau_L \propto L^z \propto L^{13/6} \approx
L^2 (1+1/6 \ln L)$.  Universality of the amplitude of a logarithmic
correction would be quite unusual and does not fit into any theoretical
framework of which we are aware.

Some open questions remain regarding the optimized variational vectors
to which this computation owes its accuracy.  Variational basis
vectors of the general form given in Eq.~(\ref{eq.psit}) are special
in the sense that all parameters enter linearly.  The method outlined
here does not require this feature and in fact it was not present in
previous computations, reported in Ref.~\onlinecite{NBdyn.prl.96}.

For most of the results presented here we used trial vectors with
linear parameters optimized by minimization of the variance of the
configurational eigenvalues.  As an apparently equivalent alternative,
the full set of symmetrized monomials in the Fourier coefficients of
the spin configuration could be chosen as basis vectors rather than
the linear combinations defined in Eq.~(\ref{eq.psit}).  With this
choice, the basis vectors would not have contained any parameters, but
employing this bigger truncated basis, the same linear parameters
would have been reintroduced by computing the matrix elements
$\hN_{ij}$ and $\hcP_{ij}$ and solving the generalized eigenvalue
problem defined by Eq.~(\ref{eq.geneig}).  In this way, we would have
obtained the coefficients for which the Rayleigh quotient is
stationary, at least if the summation over configurations could have
been done exactly.
Proceeding in this way, we could have altogether skipped the
optimization scheme based on minimization of the variance of the
configurational eigenvalues [\cf\ Eq.~(\ref{eq.chi^2}) and Section
\ref{sec.more.states}].

The obvious question is what is accomplished by the non-linear
minimization of the variance of the configurational eigenvalues.  We
do not yet have a convincing answer to this question.  On the one
hand, it is not difficult to show that the zero variance principle
holds for individual eigenstates.  That is, if an eigenvector can
represented exactly as a linear combination of the basis vectors, the
variational ($t=0$) case will already produce the exact result even if
it the eigenvalue problem contains eigenvectors that cannot be
represented exactly in the truncated basis.  This is in fact precisely
what happens for the dominant even eigenvector: this vector is
represented exactly even in the truncated basis we use, and indeed its
eigenvalue is reproduced exactly.  On the other hand, our tentative
numerical experiments show that the optimization method produces more
accurate results, which is not surprising when one considers that the
optimized basis functions give rise to much smaller truncated basis
sets.  This in turn yields a generalized eigenvalue problem involving
much smaller matrices that are numerically and statistically much more
robust.
 
A related problem with a large basis set is that the matrix $\hN$ in the
generalized eigenvalue problem of Eq.~(\ref{eq.geneig}) becomes
numerically singular.  In fact, the only way in which we were able to
obtain meaningful results at all is by performing the inversion in the
usual regularized fashion as follows.  Use the fact that $\hN$ is
symmetric and non-negative definite to write it in the form
\begin{equation}
\hN=W\,\diag(\mu_1^2,\dots,\mu_n^2)\, W^{\dagger}.
\end{equation}
Then define a regularized inverse of $\hN^{1 \over 2}$ as follows
\begin{equation}
\bar{N}^{-{1 \over 2}}=W\,\diag(\bar\mu_1^{-1},\dots,\bar\mu_n^{-1})\,
W^{\dagger},
\end{equation}
where $\bar\mu_i^{-1}=\mu_i^{-1}$ if $\mu_i$ exceeds a suitable chosen
threshold, \eg\ the square root of the machine accuracy, and
$\bar\mu_i^{-1}=0$ otherwise.  The non-vanishing eigenvalues of
$\bar{N}^{-{1 \over 2}}\hP\bar{N}^{-{1 \over 2}}$ then yield a subset
of the eigenvalues of $\hP$ that are least affected by the numerical
singularity of $\hN$.

Although further modifications of the computational procedures may lead to
additional improvements of our technique, the numerical results obtained
thus far are already quite promising and the question arises what further
applications are obvious in the field of dynamics of Monte Carlo methods.

For instance, it seems well possible to apply the present techniques
in three dimensions, and to spin-conserving Kawasaki dynamics
\cite{KD} although it is clear that the construction of trial vectors
will have to be modified. In the same context, we note that direct
application of the method used in this paper to the dynamics of
cluster algorithms \cite{SW,W,HB} is frustrated by the requirement
that one should be able to compute $\br S|\P\psiTket$ numerically
exactly. An additional problem for such dynamics from the perspective
of our approach is that the concept {\em correlation time} has to be
handled carefully in this context.

Let us demonstrate this point by means of the following thought
experiment: the application of the Wolff algorithm to the
ferromagnetic, critical Ising model.  As usual, we define the
autocorrelation times in terms of the eigenvalues of the stochastic
matrix. In order to enable a comparison with other types of dynamics,
we choose our unit of time as $L^{2d-2y_h}$ Wolff steps ($L$ is the
linear system size, $d$ the dimensionality and $y_h$ the magnetic
renormalization exponent). Since the average Wolff cluster consists of
a number of sites proportional to $L^{2 y_h-d}$, this choice
guarantees that, under equilibrium conditions, an average number of
order $L^d$ spins is processed per unit of time. 

Because of the efficiency of the Wolff algorithm, only a few units of
time are needed to generate an independent spin configuration under
practical circumstances. However, if the fully ordered
antiferromagnetic state is chosen as the initial spin configuration, a
number of Wolff steps of order $L^d$ is required to remove the the
antiferromagnetic order, \ie, its relaxation to equilibrium is
anomalously slow. A less extreme but related phenomenon is observed
under practical Wolff simulation conditions at equilibrium: from time
to time large critical fluctuations occur that bring the system in a
state of relatively large disorder and small magnetization. These
configurations are relatively long-lived. In the time-autocorrelation
function of the magnetization, this phenomenon translates into a
slower-than-exponential decay \cite{Kerl}, at least on the numerically
accessible time scale.
In the language of Eq.~(\ref{eq.auto}) such a situation follows if 
one assumes the existence of anomalously large autocorrelation times
$\tau_{Li}$ associated with anomalously small amplitudes $c_i$.
Under these circumstances we cannot exclude the
possibility that the longest relaxation time following from the Markov
matrix for the Wolff simulation of a finite system corresponds with an
extremely unlikely deviation from equilibrium. Since this kind of
fluctuations may have too low a probability to be of practical
significance, these considerations suggest the possibility that the
time needed to generate an `independent configuration' is not simply
related to the second largest eigenvalue
of the Markov matrix, but rather to some intricate average, possibly
involving the complete spectrum.

\acknowledgements 

It is our pleasure to acknowledge stimulating discussions with Bob Swendsen.
This research was supported by the (US) National Science Foundation
(NSF) through Grants DMR-9725080. This research was conducted in part
using the resources of the Cornell Theory Center, which received major
funding from NSF and New York State, with additional support from the
Advanced Research Projects Agency (ARPA), the National Center for
Research Resources at the National Institutes of Health (NIH), the IBM
Corporation, and other members of the center's Corporate Research
Institute.  This research is supported in part by the Dutch FOM
foundation (`Stichting voor Fundamenteel Onderzoek der Materie') which
is financially supported by the NWO (`Nederlandse Organisatie voor
Wetenschappelijk Onderzoek').

\newpage
\section*{Appendix A: Numerical results for the autocorrelation times}

This Appendix contains Table \ref{tab:eivas} with numerical results
for the eigenvalues of
the Markov matrix for five relaxation modes of three Ising-like models
as described in Section \ref{sec.numerical}. In order to reduce the
amount of data, results of different runs for the same model and system
size are combined into a single entry. The autocorrelation times are
directly related to these data via Eq.~(\ref{tau}).

\begin{table}[t]
\caption{ Eigenvalues of the stochastic matrix for different relaxation
modes and Ising-like models.  The first column indicates the relaxation
mode, the second one refers to the linear system size. Next follow the
eigenvalues for the three models, identified by the sign of the
next-nearest-neighbor coupling determined by $\kappa$. Each eigenvalue
is followed by the statistical error in its last decimal place between
parentheses.
Also shown are exact numerical results for system sizes up to $L=5$.
The latter entries are thought to be accurate in all decimal places given.
}
\vskip 1 ex
\begin{center}
\begin{tabular}{||c|r|lr|lr|lr||}
mode&$L$&$\kappa=1$&        &$  \kappa=2$  &        &  $\kappa=3$  &        \\
\hline
1& 2& 0.9907787946 & $(  0)$ & 0.9857022604 & $( 0)$ & 0.9795775025 & $( 0)$\\ 
1& 3& 0.9980901680 & $(  0)$ & 0.9974093851 & $( 0)$ & 0.9960331145 & $( 0)$\\ 
1& 4& 0.9994535201 & $(  0)$ & 0.9992455674 & $( 0)$ & 0.9988183100 & $( 0)$\\ 
1& 5& 0.9997956758 & $(  0)$ & 0.9997089536 & $( 0)$ & 0.9995421579 & $( 0)$\\ 
1& 5& 0.9997956748 & $(199)$ & 0.9997089585 & $(78)$ & 0.9995421589 & $(27)$\\ 
1& 6& 0.9999079148 & $( 81)$ & 0.9998657202 & $(34)$ & 0.9997887914 & $(28)$\\ 
1& 7& 0.9999527031 & $( 49)$ & 0.9999299745 & $(24)$ & 0.9998899946 & $(18)$\\ 
1& 8& 0.9999732910 & $( 35)$ & 0.9999600883 & $(17)$ & 0.9999373860 & $(14)$\\ 
1& 9& 0.9999838153 & $( 24)$ & 0.9999756656 & $(12)$ & 0.9999618670 & $(12)$\\ 
1&10& 0.9999896416 & $( 16)$ & 0.9999843583 & $(10)$ & 0.9999755105 & $( 9)$\\ 
1&11& 0.9999930699 & $( 12)$ & 0.9999895061 & $( 8)$ & 0.9999835841 & $( 7)$\\ 
1&12& 0.9999951964 & $(  9)$ & 0.9999927095 & $( 6)$ & 0.9999886009 & $( 7)$\\ 
1&13& 0.9999965694 & $(  9)$ & 0.9999947838 & $( 5)$ & 0.9999918506 & $( 6)$\\ 
1&14& 0.9999974860 & $( 14)$ & 0.9999961737 & $( 4)$ & 0.9999940233 & $( 8)$\\ 
1&15& 0.9999981168 & $(  7)$ & 0.9999971315 & $( 4)$ & 0.9999955211 & $( 6)$\\ 
1&16& 0.9999985619 & $(  6)$ & 0.9999978080 & $( 6)$ & 0.9999965798 & $( 6)$\\ 
1&17& 0.9999988856 & $(  6)$ & 0.9999982987 & $( 5)$ & 0.9999973461 & $( 6)$\\ 
1&18& 0.9999991237 & $(  5)$ & 0.9999986606 & $( 5)$ & 0.9999979110 & $( 5)$\\ 
1&19& 0.9999993003 & $(  4)$ & 0.9999989315 & $( 4)$ & 0.9999983322 & $( 4)$\\ 
1&20& 0.9999994351 & $(  4)$ & 0.9999991370 & $( 4)$ & 0.9999986546 & $( 4)$\\ 
1&21& 0.9999995400 & $(  4)$ & 0.9999992955 & $( 5)$ & 0.9999989017 & $( 4)$\\ 
2& 2& 0.926394028 & $(   0)$ & 0.833333333 & $(  0)$ & 0.801400401 & $(  0)$\\ 
2& 3& 0.974142769 & $(   0)$ & 0.952223659 & $(  0)$ & 0.938590450 & $(  0)$\\ 
2& 4& 0.990624236 & $(   0)$ & 0.983361880 & $(  0)$ & 0.976710275 & $(  0)$\\ 
2& 5& 0.995938051 & $(   0)$ & 0.993017648 & $(  0)$ & 0.989713640 & $(  0)$\\ 
2& 5& 0.995937787 & $( 498)$ & 0.993017497 & $(153)$ & 0.989713696 & $( 53)$\\ 
2& 6& 0.997991170 & $( 145)$ & 0.996638064 & $( 77)$ & 0.994916034 & $( 43)$\\ 
2& 7& 0.998906242 & $(  91)$ & 0.998205076 & $( 59)$ & 0.997249371 & $( 40)$\\ 
2& 8& 0.999359061 & $(  62)$ & 0.998962435 & $( 38)$ & 0.998398932 & $( 27)$\\ 
2& 9& 0.999601758 & $(  45)$ & 0.999361719 & $( 30)$ & 0.999011349 & $( 22)$\\ 
2&10& 0.999740631 & $(  31)$ & 0.999587240 & $( 24)$ & 0.999359261 & $( 17)$\\ 
2&11& 0.999824363 & $(  33)$ & 0.999721993 & $( 19)$ & 0.999567856 & $( 15)$\\ 
2&12& 0.999877075 & $(  20)$ & 0.999806230 & $( 17)$ & 0.999698592 & $( 14)$\\ 
\hline
\end{tabular}
\end{center}
\label{tab:eivas}
\end{table}
\newpage
\begin{table}[t]
\begin{center}
\begin{tabular}{||c|r|lr|lr|lr||}
mode&$L$&$\kappa=1$&        &$  \kappa=2$  &        &  $\kappa=3$  &        \\
\hline
2&13& 0.999911507 & $(  18)$ & 0.999861038 & $( 13)$ & 0.999783763 & $( 12)$\\ 
2&14& 0.999934850 & $(  34)$ & 0.999897876 & $( 16)$ & 0.999841006 & $( 16)$\\ 
2&15& 0.999950946 & $(  14)$ & 0.999923346 & $( 12)$ & 0.999880657 & $( 12)$\\ 
2&16& 0.999962406 & $(  24)$ & 0.999941407 & $( 10)$ & 0.999908745 & $( 12)$\\ 
2&17& 0.999970776 & $(  11)$ & 0.999954460 & $( 11)$ & 0.999929099 & $( 12)$\\ 
2&18& 0.999976911 & $(  10)$ & 0.999964107 & $( 11)$ & 0.999944100 & $(  9)$\\ 
2&19& 0.999981548 & $(   7)$ & 0.999971344 & $( 10)$ & 0.999955374 & $(  9)$\\ 
2&20& 0.999985070 & $(   6)$ & 0.999976856 & $( 15)$ & 0.999963952 & $(  8)$\\ 
2&21& 0.999987826 & $(   7)$ & 0.999981121 & $(  9)$ & 0.999970574 & $(  9)$\\ 
3& 2& 0.347072809 & $(   0)$ & 0.485702260 & $(  0)$ & 0.521667390 & $(  0)$\\ 
3& 3& 0.948416869 & $(   0)$ & 0.910335343 & $(  0)$ & 0.892928760 & $(  0)$\\ 
3& 4& 0.977929776 & $(   0)$ & 0.967112147 & $(  0)$ & 0.958932014 & $(  0)$\\ 
3& 5& 0.990525842 & $(   0)$ & 0.985750715 & $(  0)$ & 0.980972217 & $(  0)$\\ 
3& 5& 0.990520317 & $(1494)$ & 0.985750628 & $(170)$ & 0.980972191 & $( 46)$\\ 
3& 6& 0.995423912 & $( 368)$ & 0.993016576 & $(133)$ & 0.990186420 & $( 63)$\\ 
3& 7& 0.997556058 & $( 203)$ & 0.996236739 & $(108)$ & 0.994531208 & $( 41)$\\ 
3& 8& 0.998587315 & $( 109)$ & 0.997813716 & $( 68)$ & 0.996755337 & $( 35)$\\ 
3& 9& 0.999130292 & $(  79)$ & 0.998651324 & $( 48)$ & 0.997971453 & $( 33)$\\ 
3&10& 0.999437033 & $(  78)$ & 0.999126354 & $( 40)$ & 0.998674548 & $( 26)$\\ 
3&11& 0.999620464 & $(  80)$ & 0.999411001 & $( 36)$ & 0.999101009 & $( 22)$\\ 
3&12& 0.999735454 & $(  48)$ & 0.999589321 & $( 21)$ & 0.999370552 & $( 22)$\\ 
3&13& 0.999810121 & $(  50)$ & 0.999705427 & $( 19)$ & 0.999547189 & $( 23)$\\ 
3&14& 0.999860308 & $(  52)$ & 0.999783514 & $( 29)$ & 0.999666372 & $( 27)$\\ 
3&15& 0.999895086 & $(  19)$ & 0.999837459 & $( 21)$ & 0.999749207 & $( 20)$\\ 
3&16& 0.999919697 & $(  16)$ & 0.999875726 & $( 18)$ & 0.999807990 & $( 17)$\\ 
3&17& 0.999937600 & $(  18)$ & 0.999903448 & $( 15)$ & 0.999850646 & $( 17)$\\ 
3&18& 0.999950809 & $(  20)$ & 0.999923918 & $( 24)$ & 0.999882182 & $( 14)$\\ 
3&19& 0.999960665 & $(  12)$ & 0.999939274 & $( 16)$ & 0.999905880 & $( 13)$\\ 
3&20& 0.999968238 & $(  11)$ & 0.999950922 & $( 10)$ & 0.999923938 & $( 13)$\\ 
3&21& 0.999974075 & $(  10)$ & 0.999959953 & $( 16)$ & 0.999937865 & $( 13)$\\ 
4& 2&  0.21369462 & $(   0)$ &  0.16666667 & $(  0)$ &  0.37907914 & $(  0)$\\ 
4& 3&  0.86122990 & $(   0)$ &  0.84440186 & $(  0)$ &  0.82248499 & $(  0)$\\ 
4& 4&  0.96905674 & $(   0)$ &  0.94884183 & $(  0)$ &  0.93508618 & $(  0)$\\ 
4& 5&  0.98503803 & $(   0)$ &  0.97749916 & $(  0)$ &  0.97028049 & $(  0)$\\ 
4& 5&  0.98491042 & $(1831)$ &  0.97749751 & $( 76)$ &  0.97028093 & $( 65)$\\ 
4& 6&  0.99259567 & $( 170)$ &  0.98881663 & $( 35)$ &  0.98455832 & $( 19)$\\ 
\hline
\end{tabular}
\end{center}
\end{table}
\newpage
\begin{table}[t]
\begin{center}
\begin{tabular}{||c|r|lr|lr|lr||}
mode&$L$&$\kappa=1$&        &$  \kappa=2$  &        &  $\kappa=3$  &        \\
\hline
4& 7&  0.99601508 & $(  55)$ &  0.99391580 & $( 26)$ &  0.99128862 & $( 13)$\\ 
4& 8&  0.99768980 & $(  20)$ &  0.99644378 & $( 17)$ &  0.99477704 & $( 13)$\\ 
4& 9&  0.99857503 & $(  14)$ &  0.99779768 & $( 15)$ &  0.99670967 & $( 10)$\\ 
4&10&  0.99907678 & $(  12)$ &  0.99857011 & $( 14)$ &  0.99783927 & $(  8)$\\ 
4&11&  0.99937693 & $(  13)$ &  0.99903453 & $( 15)$ &  0.99852907 & $(  8)$\\ 
4&12&  0.99956541 & $(  14)$ &  0.99932578 & $( 10)$ &  0.99896776 & $(  6)$\\ 
4&13&  0.99968782 & $(   8)$ &  0.99951582 & $(  8)$ &  0.99925607 & $(  8)$\\ 
4&14&  0.99977019 & $(  12)$ &  0.99964398 & $(  7)$ &  0.99945137 & $(  6)$\\ 
4&15&  0.99982705 & $(   4)$ &  0.99973256 & $(  7)$ &  0.99958709 & $(  9)$\\ 
4&16&  0.99986788 & $(   5)$ &  0.99979523 & $(  6)$ &  0.99968380 & $(  8)$\\ 
4&17&  0.99989728 & $(   6)$ &  0.99984095 & $(  6)$ &  0.99975401 & $(  9)$\\ 
4&18&  0.99991894 & $(   3)$ &  0.99987474 & $(  6)$ &  0.99980573 & $(  6)$\\ 
4&19&  0.99993530 & $(   2)$ &  0.99989990 & $(  5)$ &  0.99984490 & $(  9)$\\ 
4&20&  0.99994758 & $(   3)$ &  0.99991903 & $(  3)$ &  0.99987466 & $(  7)$\\ 
4&21&  0.99995730 & $(   5)$ &  0.99993422 & $( 13)$ &  0.99989756 & $(  6)$\\ 
5& 3&  0.85491947 & $(   0)$ &  0.78895600 & $(  0)$ &  0.73105899 & $(  0)$\\ 
5& 4&  0.94351018 & $(   0)$ &  0.92340590 & $(  0)$ &  0.90501381 & $(  0)$\\ 
5& 5&  0.98044407 & $(   0)$ &  0.96699772 & $(  0)$ &  0.95702204 & $(  0)$\\ 
5& 5&  0.97941691 & $(7959)$ &  0.96698599 & $(529)$ &  0.95702123 & $(107)$\\ 
5& 6&  0.98907715 & $(1314)$ &  0.98341417 & $( 86)$ &  0.97770958 & $( 19)$\\ 
5& 7&  0.99396378 & $( 553)$ &  0.99088189 & $( 36)$ &  0.98733988 & $( 18)$\\ 
5& 8&  0.99648079 & $(  69)$ &  0.99463058 & $( 30)$ &  0.99233151 & $( 15)$\\ 
5& 9&  0.99782078 & $(  43)$ &  0.99665705 & $( 22)$ &  0.99512175 & $( 16)$\\ 
5&10&  0.99858562 & $(  20)$ &  0.99782136 & $( 27)$ &  0.99677069 & $( 12)$\\ 
5&11&  0.99904349 & $(  19)$ &  0.99852498 & $( 19)$ &  0.99778862 & $( 12)$\\ 
5&12&  0.99933354 & $(  19)$ &  0.99896801 & $( 14)$ &  0.99844089 & $( 12)$\\ 
5&13&  0.99952045 & $(   9)$ &  0.99925806 & $( 13)$ &  0.99887273 & $( 11)$\\ 
5&14&  0.99964737 & $(  17)$ &  0.99945408 & $( 11)$ &  0.99916587 & $( 10)$\\ 
5&15&  0.99973485 & $(  11)$ &  0.99958961 & $(  9)$ &  0.99937119 & $(  8)$\\ 
5&16&  0.99979726 & $(   9)$ &  0.99968586 & $(  7)$ &  0.99951767 & $( 12)$\\ 
5&17&  0.99984213 & $(   7)$ &  0.99975594 & $(  8)$ &  0.99962420 & $(  9)$\\ 
5&18&  0.99987534 & $(   7)$ &  0.99980723 & $(  7)$ &  0.99970320 & $( 11)$\\ 
5&19&  0.99990037 & $(   4)$ &  0.99984619 & $(  6)$ &  0.99976257 & $(  7)$\\ 
5&20&  0.99991947 & $(   4)$ &  0.99987569 & $(  5)$ &  0.99980798 & $(  7)$\\ 
5&21&  0.99993434 & $(   5)$ &  0.99989841 & $(  5)$ &  0.99984303 & $(  6)$\\ 
\hline
\end{tabular}
\end{center}
\end{table}

\newpage
\section*{Appendix B: Numerical fits involving $z$ }

This Appendix lists Table \ref{tab:zfits} containing a selection of fits
used to determine the dynamic exponent $z$. These fits were made for 
five relaxation modes of three Ising-like models. For a given number of
corrections-to-scaling $n_c$ the minimum system size must not be too
small in order to get an acceptable fit as becomes apparent on the
basis of the $\chi^2$ criterion. The Table contains a number of entries
illustrating this point. It includes data for the amplitudes $\alpha_0$
which support the hypothesis of universal amplitude ratios.  However,
statistically more accurate amplitudes (see Table \ref{tab:abest})
can be determined on the basis of the assumption $z=13/6$ which is
consistent with the results in Table \ref{tab:zfits}.

\begin{table}[t]
\caption{ Fits used for the determination of the dynamic exponent $z$. 
These fits were made for different relaxation modes ($i$, column 1) of
three distinct models identified by their value of $\kappa$ (column 2).
The third column indicates the minimum system size,
and the fourth one the number of correction terms $n_c$ that
were included in the fit. The following columns list the estimated
dynamic exponent, its statistical uncertainty (two standard deviations),
the residual $\chi^2$, the number of degrees of freedom, and the amplitude
of the $L^z$ term describing the pertinent relaxation mode.
}
\vskip 1 ex
\begin{center}
\begin{tabular}{||r|r|r|r|r|r|r|r|r||}
$i$&$\kappa$&$L\geq$&$n_c$&$z$&$\pm$ &  $\chi^2$  & $d_f$ &  $\alpha_0$   \\
\hline
1&1&  8 &   1 &   2.16352 &  0.00215 &       76.2 &    60 &       6.8322  \\
1&1&  9 &   1 &   2.16478 &  0.00306 &       71.0 &    55 &       6.8056  \\
1&1& 10 &   1 &   2.16741 &  0.00443 &       57.7 &    50 &       6.7496  \\
1&1&  5 &   2 &   2.15707 &  0.00176 &      132.5 &    74 &       6.9698  \\
1&1&  6 &   2 &   2.16405 &  0.00280 &       89.7 &    69 &       6.8206  \\
1&1&  7 &   2 &   2.16555 &  0.00452 &       86.1 &    64 &       6.7880  \\
1&1&  4 &   3 &   2.16253 &  0.00234 &      105.4 &    74 &       6.8506  \\
1&1&  5 &   3 &   2.16946 &  0.00420 &       89.2 &    73 &       6.7004  \\
1&1&  6 &   3 &   2.16870 &  0.00746 &       87.9 &    68 &       6.7173  \\
1&1&  3 &   4 &   2.16640 &  0.00271 &       95.5 &    74 &       6.7656  \\
1&1&  4 &   4 &   2.17198 &  0.00533 &       89.4 &    73 &       6.6438  \\
1&1&  5 &   4 &   2.16980 &  0.01027 &       89.2 &    72 &       6.6926  \\
1&1&  2 &   5 &   2.16815 &  0.00293 &       92.8 &    74 &       6.7270  \\
1&1&  3 &   5 &   2.17290 &  0.00605 &       89.6 &    73 &       6.6231  \\
1&1&  4 &   5 &   2.17033 &  0.01243 &       89.4 &    72 &       6.6813  \\
1&2&  5 &   1 &   2.17011 &  0.00027 &      240.5 &    77 &       4.3667  \\
1&2&  6 &   1 &   2.16791 &  0.00046 &       95.4 &    72 &       4.3931  \\
1&2&  7 &   1 &   2.16700 &  0.00072 &       83.5 &    67 &       4.4044  \\
1&2&  4 &   2 &   2.16465 &  0.00041 &       91.2 &    77 &       4.4360  \\
1&2&  5 &   2 &   2.16537 &  0.00081 &       86.9 &    76 &       4.4267  \\
1&2&  6 &   2 &   2.16559 &  0.00141 &       83.4 &    71 &       4.4238  \\
1&2&  3 &   3 &   2.16470 &  0.00050 &       91.2 &    77 &       4.4355  \\
1&2&  4 &   3 &   2.16573 &  0.00110 &       86.8 &    76 &       4.4218  \\
1&2&  5 &   3 &   2.16576 &  0.00209 &       86.8 &    75 &       4.4214  \\
1&2&  3 &   4 &   2.16594 &  0.00129 &       86.8 &    76 &       4.4188  \\
1&2&  4 &   4 &   2.16578 &  0.00262 &       86.8 &    75 &       4.4211  \\
1&2&  5 &   4 &   2.16595 &  0.00498 &       86.7 &    74 &       4.4186  \\
1&2&  2 &   5 &   2.16602 &  0.00138 &       86.8 &    76 &       4.4178  \\
1&2&  3 &   5 &   2.16602 &  0.00295 &       86.8 &    75 &       4.4178  \\
1&2&  4 &   5 &   2.16606 &  0.00597 &       86.7 &    74 &       4.4170  \\
1&3&  8 &   1 &   2.17020 &  0.00070 &       56.0 &    51 &       2.7996  \\
1&3&  9 &   1 &   2.16927 &  0.00104 &       50.1 &    47 &       2.8075  \\
1&3& 10 &   1 &   2.16882 &  0.00155 &       49.4 &    43 &       2.8114  \\
\hline
\end{tabular}
\end{center}
\label{tab:zfits}
\end{table}
\newpage
\begin{table}
\begin{center}
\begin{tabular}{||r|r|r|r|r|r|r|r|r||}
$i$&$\kappa$&$L\geq$&$n_c$&$z$&$\pm$ &  $\chi^2$  & $d_f$ &  $\alpha_0$   \\  
\hline
1&3&  4 &   2 &   2.16952 &  0.00026 &      213.9 &    63 &       2.8082  \\
1&3&  5 &   2 &   2.16674 &  0.00051 &       54.9 &    62 &       2.8306  \\
1&3&  6 &   2 &   2.16679 &  0.00085 &       53.8 &    58 &       2.8302  \\
1&3&  3 &   3 &   2.16727 &  0.00031 &       91.4 &    63 &       2.8268  \\
1&3&  4 &   3 &   2.16561 &  0.00068 &       61.5 &    62 &       2.8407  \\
1&3&  5 &   3 &   2.16695 &  0.00124 &       54.8 &    61 &       2.8288  \\
1&3&  2 &   4 &   2.16711 &  0.00034 &       87.9 &    63 &       2.8281  \\
1&3&  3 &   4 &   2.16537 &  0.00079 &       64.2 &    62 &       2.8429  \\
1&3&  4 &   4 &   2.16747 &  0.00156 &       54.3 &    61 &       2.8239  \\
1&3&  2 &   5 &   2.16528 &  0.00087 &       65.3 &    62 &       2.8437  \\
1&3&  3 &   5 &   2.16785 &  0.00177 &       54.2 &    61 &       2.8204  \\
1&3&  4 &   5 &   2.16811 &  0.00372 &       54.2 &    60 &       2.8179  \\
2&1&  8 &   1 &   2.17081 &  0.00178 &       50.6 &    30 &       0.2482  \\
2&1&  9 &   1 &   2.16728 &  0.00252 &       32.1 &    26 &       0.2510  \\
2&1& 10 &   1 &   2.16649 &  0.00356 &       25.8 &    22 &       0.2516  \\
2&1&  6 &   2 &   2.16604 &  0.00232 &       47.0 &    37 &       0.2522  \\
2&1&  7 &   2 &   2.16286 &  0.00363 &       39.2 &    33 &       0.2548  \\
2&1&  8 &   2 &   2.16072 &  0.00576 &       37.2 &    29 &       0.2566  \\
2&1&  4 &   3 &   2.16669 &  0.00194 &       62.7 &    41 &       0.2517  \\
2&1&  5 &   3 &   2.16033 &  0.00340 &       42.3 &    40 &       0.2570  \\
2&1&  6 &   3 &   2.15977 &  0.00592 &       41.7 &    36 &       0.2575  \\
2&1&  3 &   4 &   2.16209 &  0.00221 &       47.4 &    41 &       0.2556  \\
2&1&  4 &   4 &   2.15803 &  0.00429 &       42.6 &    40 &       0.2590  \\
2&1&  5 &   4 &   2.15993 &  0.00819 &       42.3 &    39 &       0.2573  \\
2&1&  2 &   5 &   2.16014 &  0.00234 &       44.4 &    41 &       0.2572  \\
2&1&  3 &   5 &   2.15743 &  0.00486 &       42.8 &    40 &       0.2595  \\
2&1&  4 &   5 &   2.16093 &  0.00999 &       42.1 &    39 &       0.2564  \\
2&2&  6 &   2 &   2.17052 &  0.00130 &       79.8 &    60 &       0.1616  \\
2&2&  7 &   2 &   2.16839 &  0.00209 &       68.2 &    56 &       0.1627  \\
2&2&  8 &   2 &   2.16899 &  0.00343 &       66.6 &    52 &       0.1624  \\
2&2&  3 &   3 &   2.17439 &  0.00047 &      209.0 &    65 &       0.1598  \\
2&2&  4 &   3 &   2.16927 &  0.00101 &       79.8 &    64 &       0.1623  \\
2&2&  5 &   3 &   2.16837 &  0.00192 &       78.6 &    63 &       0.1627  \\
2&2&  3 &   4 &   2.16819 &  0.00118 &       78.4 &    64 &       0.1629  \\
2&2&  4 &   4 &   2.16799 &  0.00240 &       78.4 &    63 &       0.1630  \\
2&2&  5 &   4 &   2.16705 &  0.00467 &       78.2 &    62 &       0.1635  \\
2&2&  2 &   5 &   2.16796 &  0.00125 &       78.4 &    64 &       0.1630  \\
2&2&  3 &   5 &   2.16793 &  0.00271 &       78.4 &    63 &       0.1630  \\
\hline
\end{tabular}
\end{center}
\end{table}
\newpage
\begin{table}[t]
\begin{center}
\begin{tabular}{||r|r|r|r|r|r|r|r|r||}
$i$&$\kappa$&$L\geq$&$n_c$&$z$&$\pm$ &  $\chi^2$  & $d_f$ &  $\alpha_0$   \\  
\hline
2&2&  4 &   5 &   2.16740 &  0.00565 &       78.4 &    62 &       0.1633  \\
2&3&  7 &   2 &   2.17079 &  0.00109 &       59.7 &    53 &       0.1037  \\
2&3&  8 &   2 &   2.16985 &  0.00182 &       56.3 &    49 &       0.1040  \\
2&3&  9 &   2 &   2.16925 &  0.00300 &       55.4 &    45 &       0.1042  \\
2&3&  5 &   3 &   2.16840 &  0.00098 &       66.0 &    60 &       0.1045  \\
2&3&  6 &   3 &   2.16689 &  0.00176 &       60.5 &    56 &       0.1050  \\
2&3&  7 &   3 &   2.16888 &  0.00319 &       58.1 &    52 &       0.1043  \\
2&3&  3 &   4 &   2.16858 &  0.00060 &       70.9 &    61 &       0.1045  \\
2&3&  4 &   4 &   2.16711 &  0.00122 &       63.3 &    60 &       0.1050  \\
2&3&  5 &   4 &   2.16646 &  0.00240 &       62.9 &    59 &       0.1052  \\
2&3&  2 &   5 &   2.16776 &  0.00064 &       65.3 &    61 &       0.1047  \\
2&3&  3 &   5 &   2.16686 &  0.00134 &       63.1 &    60 &       0.1050  \\
2&3&  4 &   5 &   2.16655 &  0.00291 &       63.1 &    59 &       0.1052  \\
3&1&  9 &   1 &   2.16225 &  0.00214 &       35.1 &    28 &       0.1206  \\
3&1& 10 &   1 &   2.16074 &  0.00310 &       32.2 &    26 &       0.1212  \\
3&1& 11 &   1 &   2.16299 &  0.00453 &       30.3 &    24 &       0.1203  \\
3&1&  7 &   2 &   2.16138 &  0.00304 &       38.9 &    31 &       0.1210  \\
3&1&  8 &   2 &   2.15949 &  0.00460 &       36.1 &    29 &       0.1217  \\
3&1&  9 &   2 &   2.16234 &  0.00746 &       35.1 &    27 &       0.1205  \\
3&1&  5 &   3 &   2.16192 &  0.00300 &       57.2 &    34 &       0.1207  \\
3&1&  6 &   3 &   2.16420 &  0.00489 &       42.2 &    32 &       0.1198  \\
3&1&  7 &   3 &   2.15988 &  0.00796 &       38.7 &    30 &       0.1216  \\
3&1&  4 &   4 &   2.16376 &  0.00376 &       56.3 &    34 &       0.1200  \\
3&1&  5 &   4 &   2.16494 &  0.00661 &       56.2 &    33 &       0.1195  \\
3&1&  6 &   4 &   2.16047 &  0.01173 &       41.7 &    31 &       0.1214  \\
3&1&  3 &   5 &   2.16536 &  0.00424 &       56.2 &    34 &       0.1193  \\
3&1&  4 &   5 &   2.16510 &  0.00795 &       56.2 &    33 &       0.1194  \\
3&1&  5 &   5 &   2.16212 &  0.01525 &       56.0 &    32 &       0.1207  \\
3&2& 10 &   2 &   2.17239 &  0.00669 &       87.6 &    45 &       0.0758  \\
3&2& 11 &   2 &   2.17316 &  0.01055 &       62.1 &    39 &       0.0756  \\
3&2& 12 &   2 &   2.17962 &  0.01806 &       51.4 &    32 &       0.0739  \\
3&2&  6 &   3 &   2.16998 &  0.00236 &       99.0 &    60 &       0.0764  \\
3&2&  7 &   3 &   2.16915 &  0.00416 &       97.6 &    56 &       0.0766  \\
3&2&  8 &   3 &   2.16681 &  0.00769 &       96.4 &    52 &       0.0772  \\
3&2&  4 &   4 &   2.17162 &  0.00171 &      105.7 &    64 &       0.0760  \\
3&2&  5 &   4 &   2.16866 &  0.00322 &      101.1 &    63 &       0.0768  \\
3&2&  6 &   4 &   2.16843 &  0.00629 &       98.7 &    59 &       0.0768  \\
\hline
\end{tabular}
\end{center}
\end{table}
\newpage
\begin{table}[t]
\begin{center}
\begin{tabular}{||r|r|r|r|r|r|r|r|r||}
$i$&$\kappa$&$L\geq$&$n_c$&$z$&$\pm$ &  $\chi^2$  & $d_f$ &  $\alpha_0$   \\  
\hline
3&2&  3 &   5 &   2.17169 &  0.00192 &      106.2 &    64 &       0.0760  \\
3&2&  4 &   5 &   2.16783 &  0.00390 &      101.1 &    63 &       0.0770  \\
3&2&  5 &   5 &   2.16842 &  0.00825 &      101.1 &    62 &       0.0768  \\
3&3&  9 &   2 &   2.17267 &  0.00227 &       56.9 &    46 &       0.0486  \\
3&3& 10 &   2 &   2.16958 &  0.00381 &       50.6 &    42 &       0.0491  \\
3&3& 11 &   2 &   2.16685 &  0.00657 &       37.1 &    34 &       0.0496  \\
3&3&  6 &   3 &   2.16940 &  0.00123 &       63.2 &    57 &       0.0492  \\
3&3&  7 &   3 &   2.16870 &  0.00235 &       61.6 &    53 &       0.0493  \\
3&3&  8 &   3 &   2.16681 &  0.00435 &       57.6 &    49 &       0.0496  \\
3&3&  4 &   4 &   2.16705 &  0.00084 &       63.8 &    61 &       0.0496  \\
3&3&  5 &   4 &   2.16680 &  0.00168 &       63.7 &    60 &       0.0496  \\
3&3&  6 &   4 &   2.16791 &  0.00345 &       62.3 &    56 &       0.0494  \\
3&3&  3 &   5 &   2.16478 &  0.00094 &       68.7 &    61 &       0.0500  \\
3&3&  4 &   5 &   2.16679 &  0.00205 &       63.8 &    60 &       0.0496  \\
3&3&  5 &   5 &   2.16801 &  0.00444 &       63.4 &    59 &       0.0494  \\
4&1&  9 &   1 &   2.16036 &  0.00283 &      113.7 &    40 &       0.0735  \\
4&1& 10 &   1 &   2.16331 &  0.00429 &      107.7 &    35 &       0.0728  \\
4&1& 11 &   1 &   2.16838 &  0.00644 &       90.3 &    27 &       0.0716  \\
4&1&  7 &   2 &   2.15902 &  0.00417 &      150.8 &    49 &       0.0738  \\
4&1&  8 &   2 &   2.16098 &  0.00652 &      129.7 &    44 &       0.0734  \\
4&1&  9 &   2 &   2.17593 &  0.01071 &      104.4 &    39 &       0.0698  \\
4&1&  4 &   3 &   2.16747 &  0.00265 &      209.3 &    54 &       0.0718  \\
4&1&  5 &   3 &   2.16523 &  0.00469 &      208.0 &    53 &       0.0723  \\
4&1&  6 &   3 &   2.17218 &  0.00747 &      152.7 &    51 &       0.0706  \\
4&1&  4 &   4 &   2.16551 &  0.00602 &      208.8 &    53 &       0.0722  \\
4&1&  5 &   4 &   2.17667 &  0.01062 &      202.1 &    52 &       0.0695  \\
4&1&  6 &   4 &   2.17489 &  0.01764 &      152.6 &    50 &       0.0700  \\
4&1&  3 &   5 &   2.15439 &  0.00666 &      225.0 &    53 &       0.0750  \\
4&1&  4 &   5 &   2.18047 &  0.01312 &      202.0 &    52 &       0.0686  \\
4&1&  5 &   5 &   2.18016 &  0.02373 &      202.0 &    51 &       0.0687  \\
4&2& 11 &   1 &   2.17894 &  0.00506 &       42.2 &    21 &       0.0448  \\
4&2& 12 &   1 &   2.17528 &  0.00743 &       40.4 &    20 &       0.0454  \\
4&2& 13 &   1 &   2.17287 &  0.01153 &       40.1 &    19 &       0.0457  \\
4&2&  7 &   2 &   2.16986 &  0.00302 &       70.4 &    33 &       0.0462  \\
4&2&  8 &   2 &   2.16652 &  0.00486 &       63.4 &    29 &       0.0467  \\
4&2&  9 &   2 &   2.17043 &  0.00817 &       52.6 &    25 &       0.0461  \\
4&2&  5 &   3 &   2.16772 &  0.00275 &       94.4 &    40 &       0.0466  \\
\hline
\end{tabular}
\end{center}
\end{table}
\newpage
\begin{table}[t]
\begin{center}
\begin{tabular}{||r|r|r|r|r|r|r|r|r||}
$i$&$\kappa$&$L\geq$&$n_c$&$z$&$\pm$ &  $\chi^2$  & $d_f$ &  $\alpha_0$   \\  
\hline
4&2&  6 &   3 &   2.16148 &  0.00473 &       77.3 &    36 &       0.0476  \\
4&2&  7 &   3 &   2.16463 &  0.00840 &       68.7 &    32 &       0.0471  \\
4&2&  4 &   4 &   2.16425 &  0.00342 &       89.2 &    40 &       0.0471  \\
4&2&  5 &   4 &   2.15912 &  0.00637 &       85.7 &    39 &       0.0480  \\
4&2&  6 &   4 &   2.16847 &  0.01258 &       75.8 &    35 &       0.0464  \\
4&2&  3 &   5 &   2.16200 &  0.00382 &       87.6 &    40 &       0.0475  \\
4&2&  4 &   5 &   2.15842 &  0.00765 &       86.4 &    39 &       0.0481  \\
4&2&  5 &   5 &   2.17347 &  0.01661 &       82.1 &    38 &       0.0456  \\
4&3&  9 &   2 &   2.18682 &  0.00544 &       82.7 &    46 &       0.0281  \\
4&3& 10 &   2 &   2.17408 &  0.00943 &       72.0 &    42 &       0.0293  \\
4&3& 11 &   2 &   2.18088 &  0.01601 &       55.7 &    34 &       0.0286  \\
4&3&  6 &   3 &   2.17547 &  0.00314 &       95.5 &    57 &       0.0292  \\
4&3&  7 &   3 &   2.17256 &  0.00556 &       93.5 &    53 &       0.0295  \\
4&3&  8 &   3 &   2.17907 &  0.01040 &       86.2 &    49 &       0.0288  \\
4&3&  4 &   4 &   2.17402 &  0.00212 &      121.9 &    61 &       0.0294  \\
4&3&  5 &   4 &   2.16574 &  0.00410 &      100.1 &    60 &       0.0302  \\
4&3&  6 &   4 &   2.17154 &  0.00818 &       94.5 &    56 &       0.0296  \\
4&3&  3 &   5 &   2.16311 &  0.00235 &      102.0 &    61 &       0.0305  \\
4&3&  4 &   5 &   2.16408 &  0.00487 &      101.8 &    60 &       0.0304  \\
4&3&  5 &   5 &   2.17429 &  0.01063 &       97.0 &    59 &       0.0293  \\
5&1&  8 &   1 &   2.17107 &  0.00225 &      135.8 &    31 &       0.0462  \\
5&1&  9 &   1 &   2.16373 &  0.00313 &       88.9 &    29 &       0.0473  \\
5&1& 10 &   1 &   2.16078 &  0.00408 &       83.4 &    27 &       0.0477  \\
5&1&  7 &   2 &   2.14959 &  0.00616 &       84.6 &    32 &       0.0496  \\
5&1&  8 &   2 &   2.14474 &  0.00709 &       76.9 &    30 &       0.0504  \\
5&1&  9 &   2 &   2.14398 &  0.01081 &       74.8 &    28 &       0.0505  \\
5&1&  5 &   3 &   2.18378 &  0.00787 &      346.5 &    35 &       0.0441  \\
5&1&  6 &   3 &   2.15418 &  0.01045 &       99.0 &    33 &       0.0488  \\
5&1&  7 &   3 &   2.13173 &  0.01418 &       77.2 &    31 &       0.0527  \\
5&1&  5 &   4 &   2.13072 &  0.01381 &      273.0 &    34 &       0.0529  \\
5&1&  6 &   4 &   2.10870 &  0.02440 &       84.1 &    32 &       0.0572  \\
5&1&  7 &   4 &   2.14044 &  0.04009 &       76.9 &    30 &       0.0511  \\
5&1&  4 &   5 &   2.08816 &  0.01550 &      272.1 &    34 &       0.0615  \\
5&1&  5 &   5 &   2.10613 &  0.03405 &      270.7 &    33 &       0.0577  \\
5&1&  6 &   5 &   2.16409 &  0.06103 &       79.6 &    31 &       0.0468  \\
5&2& 12 &   1 &   2.17736 &  0.00626 &       48.9 &    33 &       0.0293  \\
5&2& 13 &   1 &   2.17268 &  0.00931 &       36.7 &    25 &       0.0297  \\
\hline
\end{tabular}
\end{center}
\end{table}
\newpage
\begin{table}[t]
\begin{center}
\begin{tabular}{||r|r|r|r|r|r|r|r|r||}
$i$&$\kappa$&$L\geq$&$n_c$&$z$&$\pm$ &  $\chi^2$  & $d_f$ &  $\alpha_0$   \\  
\hline
5&2& 14 &   1 &   2.17800 &  0.01491 &       27.3 &    20 &       0.0292  \\
5&2&  8 &   2 &   2.17699 &  0.00432 &       74.5 &    53 &       0.0293  \\
5&2&  9 &   2 &   2.17399 &  0.00757 &       73.1 &    49 &       0.0296  \\
5&2& 10 &   2 &   2.16810 &  0.01197 &       69.1 &    45 &       0.0302  \\
5&2&  6 &   3 &   2.17310 &  0.00429 &       85.4 &    60 &       0.0297  \\
5&2&  7 &   3 &   2.16885 &  0.00765 &       79.2 &    56 &       0.0302  \\
5&2&  8 &   3 &   2.16710 &  0.01392 &       72.3 &    52 &       0.0304  \\
5&2&  4 &   4 &   2.17650 &  0.00332 &       98.2 &    64 &       0.0294  \\
5&2&  5 &   4 &   2.17080 &  0.00577 &       92.5 &    63 &       0.0300  \\
5&2&  6 &   4 &   2.16544 &  0.01138 &       83.3 &    59 &       0.0305  \\
5&2&  3 &   5 &   2.18153 &  0.00383 &      110.6 &    64 &       0.0289  \\
5&2&  4 &   5 &   2.16880 &  0.00694 &       92.0 &    63 &       0.0302  \\
5&2&  6 &   5 &   2.15753 &  0.02925 &       83.0 &    58 &       0.0314  \\
5&3& 10 &   2 &   2.18961 &  0.00778 &       75.5 &    42 &       0.0180  \\
5&3& 11 &   2 &   2.18990 &  0.01361 &       51.7 &    34 &       0.0180  \\
5&3& 12 &   2 &   2.18197 &  0.02460 &       43.3 &    27 &       0.0185  \\
5&3&  7 &   3 &   2.17721 &  0.00474 &       88.0 &    53 &       0.0188  \\
5&3&  8 &   3 &   2.17789 &  0.00861 &       84.0 &    49 &       0.0188  \\
5&3&  9 &   3 &   2.17892 &  0.01609 &       79.6 &    45 &       0.0187  \\
5&3&  5 &   4 &   2.17197 &  0.00342 &      101.0 &    60 &       0.0192  \\
5&3&  6 &   4 &   2.16525 &  0.00667 &       94.3 &    56 &       0.0197  \\
5&3&  7 &   4 &   2.17841 &  0.01385 &       88.0 &    52 &       0.0188  \\
5&3&  4 &   5 &   2.15800 &  0.00400 &       99.4 &    60 &       0.0202  \\
5&3&  5 &   5 &   2.16392 &  0.00849 &       96.8 &    59 &       0.0197  \\
5&3&  6 &   5 &   2.18450 &  0.01955 &       89.7 &    55 &       0.0184  \\
\end{tabular}
\end{center}
\end{table}
\end{document}